\documentclass[12pt]{article}

\usepackage[utf8]{inputenc}
\usepackage[british]{babel}

\usepackage{amsmath,amsthm,amssymb,amscd}
\usepackage{a4wide}
\usepackage[vcentermath]{youngtab}
\usepackage{dsfont}

\usepackage{url}

\usepackage{graphicx}
\usepackage{wrapfig}
\usepackage[bf,small]{caption}
\usepackage{enumitem}
\usepackage{tensor}

\usepackage{def}

\newlist{exc}{enumerate}{2}
\setlist[exc,1]{label={\bf\arabic*.},itemsep=.5ex,topsep=.5ex,parsep=0mm,
partopsep=0mm,resume}
\setlist[exc,2]{label=(\alph*),itemsep=0mm,topsep=0mm,parsep=0mm,
partopsep=0mm}

\usepackage{color}

\begin{document}

\thispagestyle{empty}

{\color{white} a \hfill b}
\vfill
\begin{center}
\includegraphics[width=.95\textwidth]{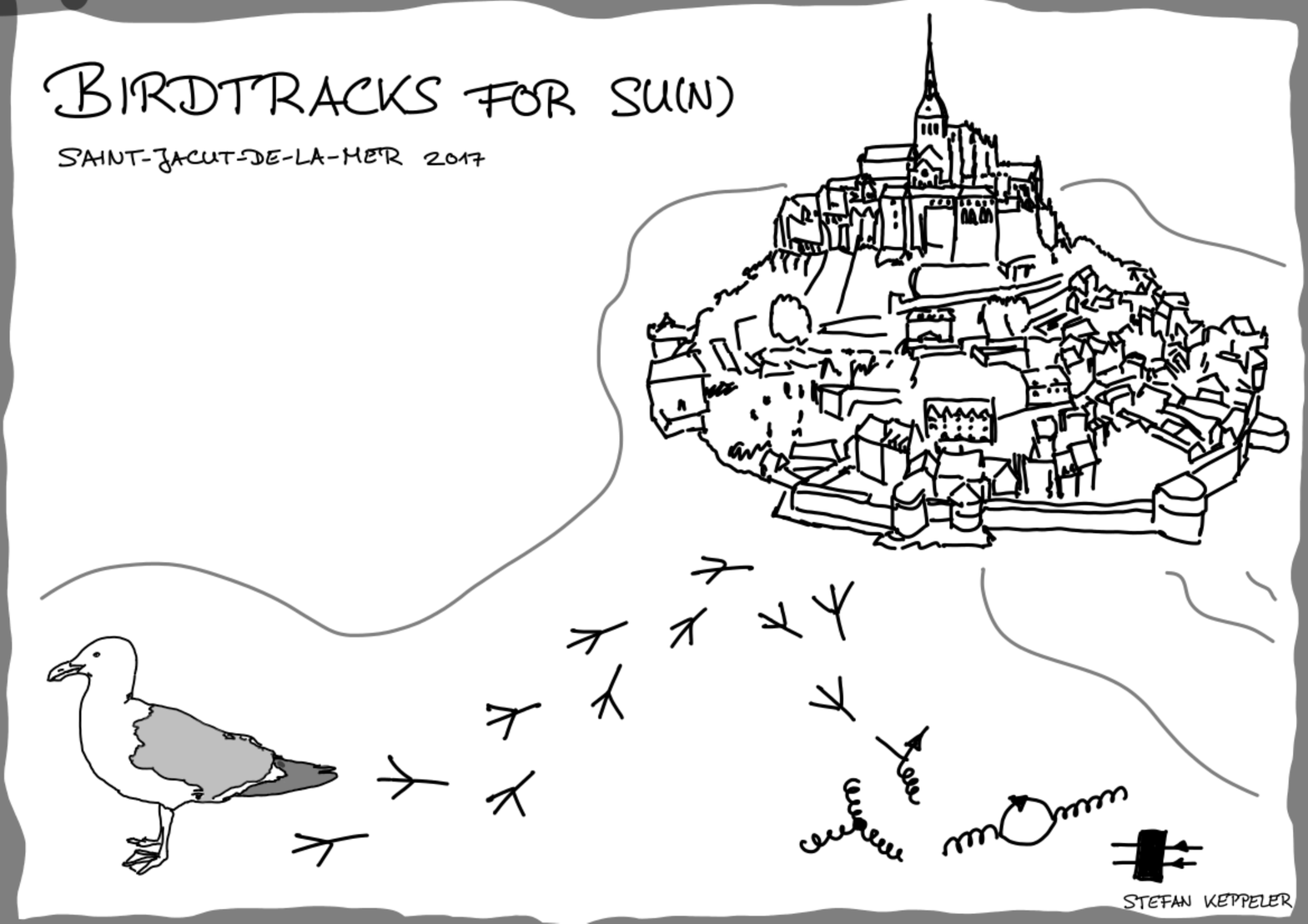}
\end{center}
\vfill
\vfill
{\color{white} c \hfill d}

\newpage


\noindent
{\large\bf Birdtracks for $\SU(N)$}\footnote{Lecture notes for the {\it QCD Master Class 2017}, \url{https://indico.cern.ch/event/547800/},\\
18--24 June 2017, Saint-Jacut-de-la-Mer, France.\\[.5ex]
\hspace*{1ex}\hfill
Please contact me if you find any errors or if you have any questions!
\hspace*{3ex}}\\[2ex]
{\bf Stefan Keppeler}\\[2ex]
Fachbereich Mathematik, Universität Tübingen\\
Auf der Morgenstelle 10, 72076 Tübingen, Germany\\
\url{stefan.keppeler@uni-tuebingen.de}
\\[2ex]

\parbox{.9\textwidth}{
  {\bf Abstract.} I gently introduce the diagrammatic birdtrack
  notation, first for vector algebra and then for permutations. After
  moving on to general tensors I review some recent results on
  Hermitian Young operators, gluon projectors, and multiplet bases for
  $\SU(N)$ colour space.}

\tableofcontents

\newpage

\section*{Introduction}
\addcontentsline{toc}{section}{Introduction}

The term {\it birdtracks} was coined by Predrag Cvitanovi\'c,
figuratively denoting the diagrammatic notation he uses in his book on
Lie groups \cite{Cvi08} -- hereinafter referred to as {\sc The
  Book}. The birdtrack notation is closely related to (abstract) index
notation. Translating back and forth between birdtracks and index
notation is achieved easily by following some simple
rules. Birdtracks, however, avoid the otherwise frequent cluttering of
indices in longer expressions. Being a notation for all sorts of
tensors, besides applications in representation theory, birdtracks are
useful in a wide range of topics, from differential geometry and
general relativity (see, e.g., \cite{Pen05}) to the classification of
semisimple Lie algebras ({\it exceptional magic} in {\sc The
  Book}). More details (and references) can, e.g., be found in
Sec.~4.9 of {\sc The Book}.

These notes were prepared for an 8~hour course aimed at graduate
students at the {\sc QCD Master Class 2017} held from 18 to 24 June
2017 in Saint-Jacut-de-la-Mer, France. I introduce the birdtrack
notation twice in well-known areas, first for vector algebra in
Sec.~\ref{sec:vectors} and later for permutations in
Sec.~\ref{sec:permutations}. The notation is extended and adapted to
$\SU(N)$ tensors in Sec.~\ref{sec:tensors}. In
Sec.~\ref{sec:color_space} I demonstrate the usefulness of birdtracks
for calculations in quantum chromodynamics (QCD). In particular, I
discuss the construction of bases for QCD colour space. The intention
of these notes is thus twofold:
Sections.~\ref{sec:vectors}-\ref{sec:permutations} contain a gentle
introduction to the birdtack notation, whereas
Sec.~\ref{sec:color_space} illustrates the use of birdtracks for QCD
colour structure.

If you would like to learn more about different flavours of
birdtracks, their history and their uses, then jump to
Sec.~\ref{sec:resources}. First, however, I'd rather you joined me on
a little journey into the world of birdtracks, starting with basic
vector algebra in the following section.


\section{Vector algebra}
\label{sec:vectors} 

We begin with vectors $\vec{a},\vec{b},\vec{c},\ldots \in \R^3$,
scalar and cross products, review the index notation and introduce the
diagrammatic birdtrack notation as illustrated in the following table. 
\vspace{1ex}
\begin{center}
\begin{tabular}{l|c|l}
& index notation & birdtrack notation\\\hline
vector $\vec{a}$ & $a_j$ & $\birdtrack{10ex}{vector_a}$ \\\hline
scalar product $\vec{a}\cdot\vec{b}$ & $a_j b_j$ &  $\birdtrack{14ex}{scalar_prod_ab}$\\\hline
cross product $\vec{a}\times\vec{b}$ & $\varepsilon_{jk\ell}a_jb_k$ & $\birdtrack{16ex}{cross_prod_ab}$
\end{tabular}
\end{center}
\vspace{1ex}
More precisely, translating back and forth between index notation and
birdtracks is achieved by assigning indices to external lines,
\begin{equation}
  a_j = \birdtrack{10ex}{vector_aj} \, .
\end{equation}
Index contractions (we always sum over repeated indices) correspond to
joining lines,
\begin{equation}
  a_j b_j = \birdtrack{14ex}{scalar_prod_ab} \, .
\end{equation}
Consequently, an isolated line is a Kronecker-$\delta$, 
\begin{equation}
  \delta_{jk} = \raisebox{1ex}{\birdtrack{8ex}{delta_jk}} \, , 
\end{equation}
(contracting with a Kronecker-$\delta$ corresponds to extending a
line). For the totally anti-symmetric $\varepsilon$,
\begin{equation}
\begin{split}
  \varepsilon_{123} = \varepsilon_{231} = \varepsilon_{312} &= 1\\
  \varepsilon_{213} = \varepsilon_{132} = \varepsilon_{321} &= -1\\
  \varepsilon_{jk\ell} &= 0 \quad \text{if at least two indices have the same value,}
\end{split}
\end{equation}
we write a vertex, 
\begin{equation}
  \varepsilon_{jk\ell} = \birdtrack{12ex}{epsilon_jkl} \, , 
\end{equation}
thereby agreeing to read off indices in counter-clockwise order.  

When not assigning indices -- which is what we want to do most of the
time -- the position where an external line ends determines which
lines have to be identified in equations. For instance, the
anti-symmetry of $\varepsilon$ in the first two indices is expressed
as
\begin{equation}
\label{eq:epsilon_twist}
  \birdtrack{12ex}{epsilon_twist} = - \birdtrack{12ex}{epsilon} \ .
\end{equation}

Now we can use this notation to write components of cross products, 
\begin{equation}
 \varepsilon_{jk\ell}a_jb_k = \birdtrack{16ex}{cross_prod_abl} \, , 
\end{equation}
where we omit labelling lines with indices over which we sum
anyway. Equivalently, omitting indices altogether,
\begin{equation}
 \vec{a} \times \vec{b} = \birdtrack{16ex}{cross_prod_ab} \ .
\end{equation}

Let's study the following diagram, 
\begin{equation}
\label{eq:epsilonepsilon}
  \birdtrack{18ex}{epsilonepsilon} \ .
\end{equation}
Assigning indices for a moment, 
\begin{equation}
  \birdtrack{18ex}{epsilonepsilon_ijkl} \, .
\end{equation}
we see that the only non-zero terms have 
\begin{center}
$i=k$ and $j=\ell$, or\\
$i=\ell$ and $j=k$,
\end{center}
since otherwise we cannot satisfy $i \neq j \neq m \neq i$ and
$m \neq \ell \neq k \neq m$. Thus, \eqref{eq:epsilonepsilon} is a
linear combination of the diagrams
\begin{equation}
  \birdtrack{12ex}{lineline} 
  \qquad \text{and} \qquad  
  \birdtrack{12ex}{crossed_lines} \ .
\end{equation}
Closer inspection shows
\begin{equation}
\label{eq:epseps_identity}
  \birdtrack{18ex}{epsilonepsilon} 
  = \birdtrack{12ex}{crossed_lines} - \birdtrack{12ex}{lineline} \ ,
\end{equation}
which is nothing but the birdtrack version of the well-known identity
\begin{equation}
\label{eq:epseps_identity_indices}
  \varepsilon_{ijm}\varepsilon_{\ell km} 
  = \delta_{i\ell}\delta_{jk} - \delta_{ik}\delta_{j\ell} \, .
\end{equation}
(Of course, we could have also obtained Eq.~\eqref{eq:epseps_identity} by
translating Eq.~\eqref{eq:epseps_identity_indices} into birdtrack notation
instead of deriving the identity within birdtrack notation.)

Equation~\eqref{eq:epseps_identity} can be used in order to derive
identities for double cross products and similar formulas from vector
algebra which are notoriously difficult to remember. For instance, 
\begin{equation}
\begin{split}
  (\vec{a}\times\vec{b})\times\vec{c}
  &= \birdtrack{26ex}{axbxc} \\[1ex]
  &= \birdtrack{20ex}{_ac_b} - \birdtrack{20ex}{_bc_a} 
  = (\vec{a}\cdot\vec{c})\vec{b} - (\vec{b}\cdot\vec{c})\vec{a} \, .
\end{split}
\end{equation}

\goodbreak

\vspace{2ex}\noindent{\bf Exercises:} 
\begin{exc}
\item Derive a similar identity for
  $(\vec{a}\times\vec{b}\,)\cdot(\vec{c}\times\vec{d}\,)$.
\item Show that
  $\big((\vec{a}\times\vec{b}\,)\times\vec{c}\big)\times\vec{d} =
  \big((\vec{a}\times\vec{b}\,)\cdot\vec{d}\,\big)\vec{c} -
  (\vec{a}\times\vec{b}\,)(\vec{c}\cdot\vec{d}\,)$.
\end{exc}
\vspace{2ex}

In birdtrack notation, it is immediately manifest that the triple product, 
\begin{equation}
  (\vec{a}\times\vec{b})\cdot\vec{c} = \birdtrack{20ex}{triple_abc} \, , 
\end{equation}
is invariant under cyclic permutations of the three vectors. 

%

Taking Eq.~\eqref{eq:epseps_identity} and joining the upper left to
the upper right line and also the lower left to the lower right line
yields
\begin{equation}
\label{eq:epseps_norm}
  \birdtrack{12ex}{eps_loop} 
  = \birdtrack{5.5ex}{fig_eight} - \birdtrack{10.5ex}{two_loops}
  = 3 - 3 \cdot 3 = -6 \, , 
\end{equation}
where we have used that each loop contributes a factor of $\delta_{jj}=3$. 

\vspace{2ex}\noindent{\bf Exercise:} 
\begin{exc}
\item Evaluate 
\begin{equation}
  \birdtrack{24ex}{epseps_line} \, .
\end{equation}
\end{exc}
\vspace{2ex}

Rotating Eq.~\eqref{eq:epseps_identity} by $90^\circ$ we obtain the
equivalent identity
\begin{equation}
  \birdtrack{12ex}{eps_on_eps} 
  = \birdtrack{12ex}{crossed_lines} - \birdtrack{12ex}{trace} \, .
\end{equation}
  
\vspace{2ex}\noindent{\bf Exercise:} 
\begin{exc}
\item Evaluate 
\begin{equation}
  \birdtrack{20ex}{eps_box} \, .
\end{equation}
\end{exc}
\vspace{2ex}

Finally, we want to study 
\begin{equation}
  \birdtrack{16ex}{epseps_free} \, .
\end{equation}
Imagine for a moment assigning indices to the lines, then the diagram
is non-zero only if the value of each index on the left matches the
value of exactly one index on the right, i.e. 
\begin{equation}
\label{eq:epseps_free}
  \birdtrack{14ex}{epseps_free}
  = A \birdtrack{7ex}{S3_id} + B \birdtrack{8ex}{S3__12_} 
    + C \birdtrack{8ex}{S3__23_} + D \birdtrack{8ex}{S3__13_} 
    + E \birdtrack{8ex}{S3__123_} + F \birdtrack{8ex}{S3__132_}
\end{equation}
with some constants $A$ to $E$. Intertwining, say the first two lines
on the right, introduces a factor of $(-1)$ on the l.h.s.\ of the
equation, see Eq.~\eqref{eq:epsilon_twist}, whereas on the r.h.s.\ the
roles of the terms interchange. Together this implies
\begin{equation}
  B=-A \, , \quad F=-C \quad \text{and} \quad E=-D \, .
\end{equation}
Similarly, by intertwining the lower two lines, we find 
\begin{equation}
  C=-A \, , \quad E=-B \quad \text{and} \quad F=-D \, , 
\end{equation}
and thus 
\begin{equation}
\label{eq:epseps_free_prop_A3}
  \birdtrack{14ex}{epseps_free}
  = A \left( \birdtrack{7ex}{S3_id} - \birdtrack{8ex}{S3__12_} 
    - \birdtrack{8ex}{S3__23_} - \birdtrack{8ex}{S3__13_} 
    + \birdtrack{8ex}{S3__123_} + \birdtrack{8ex}{S3__132_} \right) \, .
\end{equation}
The factor $A$ can, e.g., be determined by joining all lines on the
left to those on the right (first to first etc.): On the l.h.s.\ we
obtain $-6$, see Eq.~\eqref{eq:epseps_norm}, and hence
\begin{equation}
\begin{split}
  -6 
  &= A \left( \birdtrack{10ex}{S3_id_tr} - \birdtrack{10ex}{S3__12__tr} 
     - \birdtrack{10ex}{S3__23__tr} - \birdtrack{10ex}{S3__13__tr} 
     + \birdtrack{10ex}{S3__123__tr} + \birdtrack{10ex}{S3__132__tr} \right)
  \\[1.5ex]
  &= A \, ( 27 - 9 - 9 - 9 + 3 + 3)\\
  &= 6A 
  \qquad\qquad\qquad\qquad\qquad\qquad\qquad \Leftrightarrow \quad 
  A=-1 \, .
\end{split}
\end{equation}
Later, we will denote anti-symmetrisation of a couple of lines by a
solid bar over these lines, normalised by the number of terms, i.e.
\begin{equation}
\label{eq:A3}
  \birdtrack{11ex}{A3}
  = \frac{1}{3!} \left( \birdtrack{7ex}{S3_id} - \birdtrack{8ex}{S3__12_} 
    - \birdtrack{8ex}{S3__23_} - \birdtrack{8ex}{S3__13_} 
    + \birdtrack{8ex}{S3__123_} + \birdtrack{8ex}{S3__132_} \right) \, , 
\end{equation}
and likewise for symmetrisation using an open bar. With this notation
we can rewrite our result as
\begin{equation}
  \birdtrack{14ex}{epseps_free} = -6 \, \birdtrack{11ex}{A3} \ .
\end{equation}


\section{Birdtracks for $\boldsymbol{\SU(N)}$ tensors}
\label{sec:tensors}

In Sec.~\ref{sec:vectors} we have encountered a simple rule for
translating expressions from index notation to birdtrack notation:
Draw some vertex, box or blob, possibly with a name inside and attach
a line for each index. 

In later sections we will also study quantities that have several
different types of indices, an therefore we will use different types
of lines. In particular, we are interested in the following
situation. Let $V$ be a finite dimensional (complex) vector space, say
$\dim V = N$ (i.e.\ $V\cong\C^N$) and let $\overline{V}$ be its dual,
i.e.\ the space of all linear maps $V\to\C$. In index notation we
denote components\footnote{in abstract index notation also the vector
  itself} of $v\in V$ by $v^j$, with an upper index
$j=1,\ldots,N$. Components of $u\in\overline{V}$ are in turn denoted
by $u_j$, with a lower index $j=1,\ldots,N$. In order to distinguish
the two kinds of indices we write
\begin{equation}
  v^j = \birdtrack{10ex}{vector_vj}
  \qquad \text{and} \qquad
  u_k = \birdtrack{10ex}{covector_uk} \, , 
\end{equation}
i.e., arrows point away from upper indices and towards lower
indices. Since we can only contract upper indices with lower indices,
in birdtrack notation we can only connect lines whose arrows point in
the same direction, e.g.
\begin{equation}
  u(v) = u_j v^j = \birdtrack{14ex}{contraction_uv} \, , 
\end{equation}
and, consequently,
$\delta\indices{^j_k}=\raisebox{.5ex}{\birdtrack{8ex}{quark_line_jk}}$.

If $V$ carries a representation $\Gamma$ of a Lie group $G$, i.e.\
$\Gamma:G\to\GL(V)$, then $\overline{V}$ naturally carries the
contragredient representation, for which the representation matrices
are the transposes of the inverses. If the representation $\Gamma$ is
unitary, then the inverse transpose is the complex conjugate.

We are particularly interested in the case where $V=\C^N$ and where
$\Gamma$ is the defining (or fundamental) representation of
$G=\SU(N)$. Then $\overline{V}$ carries the complex conjugate of the
defining representation. Now complex conjugation of a diagram is
achieved by reversing all arrows.

Another important representation is the adjoint representation, the
representation of a Lie group $G$ on its own Lie algebra
$\mathfrak{g}$. For $G=\SU(N)$, the Lie algebra $\mathfrak{g}=\su(N)$
consists of the traceless Hermitian $N\times N$ matrices. Elements of
a basis of the Lie algebra are called generators, and $\su(N)$ is a
real vector space of dimension $N^2{-}1$. Since it is a real vector
space we do not distinguish upper and lower indices, and in birdtrack
notation we introduce a new type of line without
arrow.\footnote{However, in order to be able to take tensor products
  of $V$, $\overline{V}$ and the carrier space of the adjoint
  representation we complexify $\su(N)$ to $A\cong\C^{N^2-1}$.} In
particular, we denote generators $t^a \in \su(N)$ as \vspace{-2ex}
\begin{equation}
  (t^a)\indices{^j_k} = \raisebox{2ex}{\birdtrack{14ex}{generator}} \, ,
\end{equation}
i.e.\ we write no vertex symbol, box or blob. 

Having applications in quantum chromodynamics (QCD) in mind, we also
refer to upper and lower indices $j,k,\ell\in\{1,\ldots,N\}$ as quark
and anti-quark indices, to lines with arrows as (anti-)quark lines, to
indices $a,b,c\in\{1,\ldots,N^2-1\}$ as gluon indices, and to curly
lines as gluon lines.

Cvitanovi\'c \cite{Cvi08} draws gluon lines as thin straight
lines instead of curly lines, and in handwritten notes it is often
convenient to use wiggly lines. 

Vanishing of the trace of the generators, $\utr t^a = (t^a)\indices{^j_j}=0$,
is expressed as
\begin{equation}
  \birdtrack{14ex}{trta} = 0 
\end{equation}
in birdtrack notation. 

The Lie bracket (commutator) is given by 
\begin{equation}
\label{eq:Lie_bracket}
  [t^a,t^b] 
  = t^a t^b - t^b t^a 
  = \ui f^{abc} t^c \, , 
\end{equation}
with the totally anti-symmetric structure constants $f^{abc}$. The
latter we denote by a vertex,
\begin{equation}
  \ui f^{abc} = \birdtrack{14ex}{structure_constants} \, , 
\end{equation}
reading off indices in anti-clockwise order. 

Written in components Eq.~\eqref{eq:Lie_bracket} reads
\begin{equation}
  (t^a)\indices{^j_\ell} (t^b)\indices{^\ell_k} 
  - (t^b)\indices{^j_\ell} (t^a)\indices{^\ell_k}
  = \ui f^{abc} (t^c)\indices{^j_k} \, , 
\end{equation}
whereas in birdtrack notation we write
\begin{equation}
\label{eq:Lie_bracket_birdtrack}
  \birdtrack{18ex}{tatb} - \birdtrack{18ex}{tbta} 
  = \birdtrack{16ex}{fabctc} \,  .
\end{equation}
It is convenient to normalise the generators as follows, 
\begin{equation}
  \underset{\text{\bf matrix notation}\strut}
  {\utr (t^a t^b) = T_R \, \delta^{ab}}
  \quad \Leftrightarrow \quad
  \underset{\text{\bf index notation}\strut}
  {(t^a)\indices{^j_k} (t^b)\indices{^k_j} = T_R \, \delta^{ab}}
  \quad \Leftrightarrow \quad
  \underset{\text{\bf birdtrack notation}\strut}
  {\birdtrack{20ex}{trtatb} = T_R \birdtrack{8ex}{gluon_line}} \, ,
\end{equation}
where $T_R$ is an arbitrary normalisation constant;
$T_R=\frac{1}{2},\,1$ or $2$ are common choices. \goodbreak\noindent
Multiplying the Lie bracket \eqref{eq:Lie_bracket_birdtrack} with
another generator and taking the trace we obtain
\begin{equation}
  \birdtrack{16ex}{trtatbtc} - \ \raisebox{-.2ex}{\birdtrack{12ex}{trtbtatc}}
  \ \ = \raisebox{-.3ex}{\birdtrack{11ex}{trftt}} \, , 
\end{equation}
which can be rewritten as 
\begin{equation}
\label{eq:f_as_3loops}
  \birdtrack{16ex}{trtatbtc} - \ \birdtrack{16ex}{trtbtatc_untwist} 
  = \ \ T_R \birdtrack{11ex}{f_rotate} \, .
\end{equation}

\vspace{2ex}\noindent{\bf Exercises:} 
\begin{exc}
\item Re-derive Eq.~\eqref{eq:f_as_3loops} in matrix or index notation.
\item {\bf Decomposition of $\boldsymbol{\overline{V} \otimes V}$.}
  \addcontentsline{toc}{subsection}{Exercise: Decomposition of
    $\overline{V} \otimes V$}
  \label{ex:VbarV}
  \\
  We define the linear map
  $P_A:\overline{V} \otimes V \to \overline{V} \otimes V$ by
\begin{equation}
  \label{eq:P_A_def}
  (P_A)\indices{^j_k^\ell_m} 
  = \frac{1}{T_R} \, (t^a)\indices{^j_k}(t^a)\indices{^\ell_m} \, .
\end{equation}
\begin{exc}
\item Write $P_A$ in birdtrack notation and verify that $(P_A)^2=P_A$ (in birdtracks!).
\end{exc}
We further define the linear map
  $P_\bullet:\overline{V} \otimes V \to \overline{V} \otimes V$ by
\begin{equation}
  \label{eq:P_bullet_def}
  P_\bullet = C \, \birdtrack{12ex}{singlet} \, .
\end{equation}
\begin{exc}[resume]
\item Fix $C>0$ such that $P_\bullet^2 = P_\bullet$. 
\end{exc}
Now $P_\bullet$ and $P_A$ are projection operators. We find the
dimensions of their images (the subspaces onto which they project) by
taking the trace.
\begin{exc}[resume]
\item Determine $\utr P_\bullet$ and $\utr P_A$ (in birdtracks!).  
\item Calculate $P_\bullet P_A$ and $P_A P_\bullet$ (in birdtracks!).
\end{exc}
Apparently, $P_\bullet$ and $P_A$ project onto mutually
transversal\footnote{i.e.\ $\im P_\bullet \subset \ker P_A$ and
  $\im P_A \subset \ker P_\bullet$} subspaces. From
$\utr P_\bullet + \utr P_A = N^2$ we conclude that
$P_\bullet + P_A = \eins_{\overline{V}\otimes V}$, in birdtracks
\begin{equation}
\label{eq:decomposeVbarV}
  \birdtrack{12ex}{idVbarV} 
  = \ \frac{1}{N} \birdtrack{12ex}{singlet} 
    \ + \ \frac{1}{T_R} \birdtrack{14ex}{octet} \, .
\end{equation}
\end{exc}
\vspace{2ex}

Equation~\eqref{eq:decomposeVbarV} can be rearranged to yield the
important identity (sometimes referred to as Fierz identity)
\begin{equation}
\label{eq:remove_gluon}
  \birdtrack{14ex}{octet} 
  = T_R \birdtrack{12ex}{idVbarV} 
    - \frac{T_R}{N} \birdtrack{12ex}{singlet} \, , 
\end{equation}
which can be used in order to remove internal gluon lines from any
diagram. For instance, 
\begin{equation}
\label{eq:C_F}
  \raisebox{2ex}{\birdtrack{20ex}{reabsorb_gluon}}
  = T_R \raisebox{.2ex}{\birdtrack{20ex}{reabsorb_gluon_02}}
    - \frac{T_R}{N} \birdtrack{9ex}{reabsorb_gluon_03}
  = T_R \frac{N^2-1}{N} \birdtrack{9ex}{reabsorb_gluon_03} \, .
\end{equation}
The prefactor $C_F:=T_R \frac{N^2-1}{N}$ is known as the (quadratic)
Casimir operator in the defining/fundamental representation.

\goodbreak 
\vspace{2ex}\noindent{\bf Exercises:}
\begin{exc}[itemsep=1ex]
\item Evaluate 
\begin{equation}
  \birdtrack{22ex}{2f_loop}
\end{equation}
\goodbreak by fist replacing the structure constants with quark loops
according to Eq.~\eqref{eq:f_as_3loops} and then removing all internal
gluon lines by means of Eq.~\eqref{eq:remove_gluon}. (You should
obtain $2 T_R N \birdtrack{7ex}{gluon_line}$; the prefactor
$C_A:=2 T_R N$ is known as the (quadratic) Casimir operator in the
adjoint representation.)
\item We would expect Eq.~\eqref{eq:Lie_bracket_birdtrack} to also
  hold with the quark line $\birdtrack{7ex}{quark_line}$ replaced by a
  gluon line $\birdtrack{7ex}{gluon_line}$ (and the quark-gluon
  vertices/generators
  $\birdtrack{7ex}{generator_no_indices}$ replaced by
  triple-gluon vertices/structure constants
  $\birdtrack{7ex}{structure_constants_no_indices_02}$) -- the
  resulting equation being known as Jacobi identity. Prove the Jacobi
  identity in birdtracks as follows:
\begin{exc}
\item Replace the vertices in $\birdtrack{7ex}{fgf}$ by quark loops
  according to Eq.~\eqref{eq:f_as_3loops}.\\ Then remove all internal
  gluon lines by using Eq.~\eqref{eq:remove_gluon}.
\item Rotate the result of (a) by 90$^\circ$.
\item Cross the two lower lines of your result of (b).  
\item Subtract your result of (c) from your result of (b) and compare
  with (a).\\ This should conclude the proof of the Jacobi relation.
\end{exc}
\end{exc}
\vspace{2ex}

The transpose of a birdtrack diagram is obtained by mirroring the
diagram across a vertical line, e.g.\
\begin{equation}
  (P_\bullet)^T = C \birdtrack{12ex}{singletT} \, .
\end{equation}
Hermitian conjugation, denoted by a dagger, corresponds to transposing
and taking the complex conjugate; recall that complex conjugation
amounts to reversing all arrows.

\vspace{2ex}\noindent{\bf Exercise:}
\begin{exc}
\item Verify that $(P_A)^\dag=P_A$ and $(P_\bullet)^\dag=P_\bullet$,
  with $P_A$ and $P_\bullet$ as defined in Eqs.~\eqref{eq:P_A_def} and
  \eqref{eq:P_bullet_def}.
\end{exc} 
\vspace{2ex}

\begin{wrapfigure}{r}{.3\textwidth}
\centering
\vspace{-2ex}
\includegraphics[width=.3\textwidth]{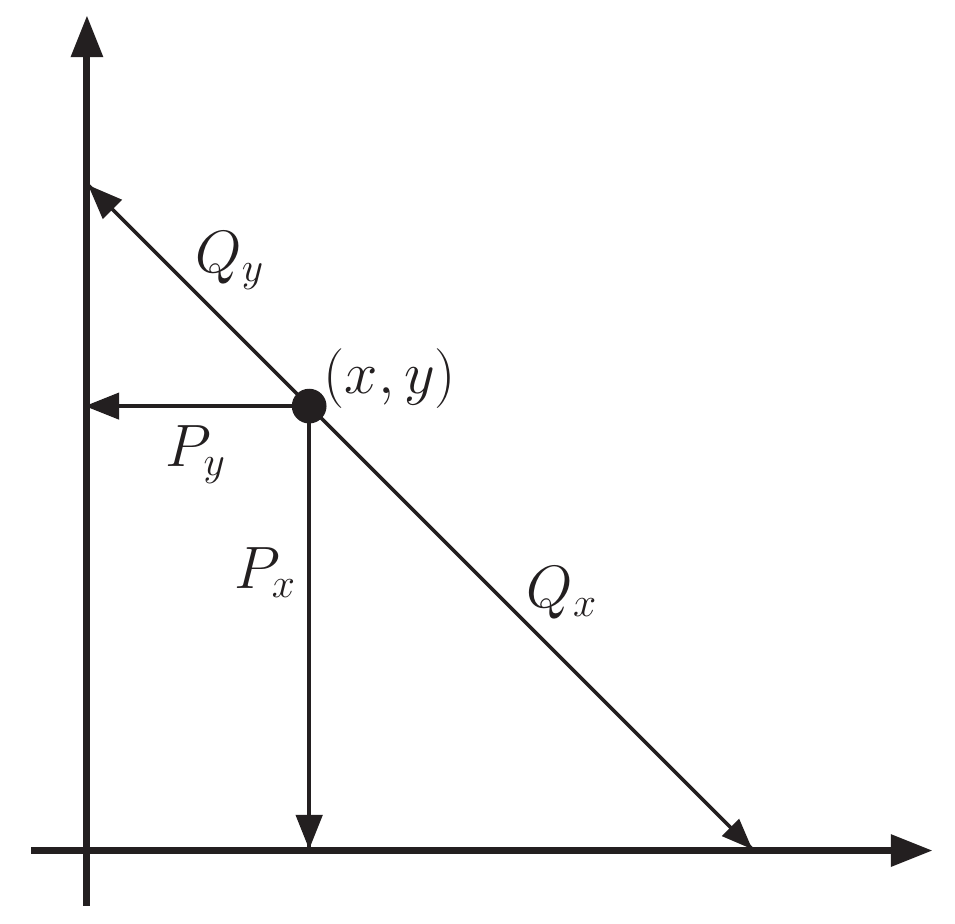}
\vspace{-7ex}
\end{wrapfigure}
Recall that Hermitian projection operators project orthogonally. Let
us illustrate this statement for projections from the $xy$-plane to
the coordinate axes. We can, e.g., project a point
$\left(\begin{smallmatrix} x \\ y \end{smallmatrix}\right)$ to the
$x$-axis or to the $y$-axis by applying the operators
\begin{equation}
P_x = \begin{pmatrix} 1 & 0 \\ 0 & 0 \end{pmatrix}
\qquad \text{or} \qquad 
P_y = \begin{pmatrix} 0 & 0 \\ 0 & 1 \end{pmatrix} \, ,
\end{equation}
respectively. $P_x$ and $P_y$ are projectors, since
$P_x^2=P_x$ and $P_y^2=P_y$, their images being the $x$- and the
$y$-axis, respectively. However, we can also project to the coordinate
axes in many other ways. For instance,
\begin{equation}
Q_x = \begin{pmatrix} 1 & 1 \\ 0 & 0 \end{pmatrix}
\qquad \text{or} \qquad 
Q_y = \begin{pmatrix} 0 & 0 \\ 1 & 1 \end{pmatrix} \, ,
\end{equation}
also satisfy and $Q_x^2=Q_x$ and $Q_y^2=Q_y$, as well as
\begin{equation}
\begin{split}
  \im Q_x &= \im P_x = \{ (x,y) \in \R^2 \, | \, y=0 \}
  \quad \text{and} \\
  \im Q_y &= \im P_y = \{ (x,y) \in \R^2 \, | \, x=0 \} \, .
\end{split}
\end{equation}
A projection operator projects onto its image along (lines parallel
to) its kernel, see figure. The kernels are
\begin{equation}
\begin{split}
  \ker P_x &= \{ (x,y) \in \R^2 \, | \, x=0 \} \, , \\
  \ker P_y &= \{ (x,y) \in \R^2 \, | \, y=0 \} \, , \\
  \ker Q_x &= \{ (x,y) \in \R^2 \, | \, y=-x \} = \ker Q_y \, .
\end{split}
\end{equation}
If a linear operator is Hermitian then its image is orthogonal to its
kernel. We have $P_x^\dag=P_x$ and $P_y^\dag=P_y$, and indeed
$\im P_x \perp \ker P_x$ and $\im P_y \perp \ker P_y$.  Moreover,
since $\im P_x$ and $\im P_y$ intersect only at the origin, we have
$P_x P_y = 0 = P_y P_x$. However, for the non-Hermitian projectors
$Q_x$ and $Q_y$ we have $Q_x Q_y \neq 0 \neq Q_y Q_x$, although their
images also intersect only at the origin.


\section{Permutations and the symmetric group}
\label{sec:permutations}

Inspecting once more Eqs.~\eqref{eq:epseps_free},
\eqref{eq:epseps_free_prop_A3} and \eqref{eq:A3} we can come up with a
different meaning for the diagrams on the right-hand side. Apparently,
we have just found a notation for permutations of 3 objects, i.e.\
elements of the symmetric group $S_3$. (We denote by $S_n$ the group
of all permutations of $n$ objects, the group multiplication being
composition of mappings.) To this end read the diagrams on the r.h.s.\
of Eq.~\eqref{eq:epseps_free}, \eqref{eq:epseps_free_prop_A3} or
\eqref{eq:A3} as mapping ends of lines from right to left. Recalling
two other standard notations for permutations we have, e.g.,
\begin{equation}
  \pi
  = \underset{\text{\bf two-line notation}}
             {\begin{pmatrix} 1 & 2 & 3 \\ 3 & 1 & 2 \end{pmatrix}}
  = \underset{\text{\bf cycle notation}\strut}{(132)}
  = \underset{\text{\bf birdtrack notation}}{\birdtrack{8ex}{S3__132_}} 
  \hspace*{-2ex} , 
\end{equation}
which all mean $\pi(1)=3$, $\pi(2)=1$, and $\pi(3)=2$. Composition
with a second permutation, e.g.\ 
\begin{equation}
  \sigma 
  = \begin{pmatrix} 1 & 2 & 3 \\ 2 & 1 & 3 \end{pmatrix}
  = (12) 
  = \birdtrack{8ex}{S3__12_} \ , 
\end{equation}
can be determined in several ways. Say, we are interested in
$\pi\circ\sigma$, we can
\begin{itemize}
\item determine individual elements
\begin{equation}
\begin{split}
  (\pi \circ \sigma)(1) &= \pi(\sigma(1)) = \pi(2) = 1\\
  (\pi \circ \sigma)(2) &= \pi(\sigma(2)) = \pi(1) = 3\\
  (\pi \circ \sigma)(3) &= \pi(\sigma(3)) = \pi(3) = 2 \, , 
\end{split}
\end{equation}
\item multiply cycles (recall that every permutation is a product of disjoint cycles)
\begin{equation}
  \pi \circ \sigma 
  = \underset{\text{\bf omit `$\boldsymbol{\circ}$'}}{(132)(12)} 
  \underset{\boldsymbol{(*)}}{=} (1)(23)
  = \underset{\text{\bf omit one-cycles}}{(23)} 
\end{equation}
{\scriptsize\bf $\boldsymbol{(*)}$ 
Write `$\boldsymbol{(1}$', where is it mapped? Thereby read from right to left.
\\[-1ex]
\phantom{$\boldsymbol{(*)}$} Continue till you'd return to $1$, then `$)$'\\[-1ex]
\phantom{$\boldsymbol{(*)}$} Repeat starting with first number not used so far.}\\
or
\item compose diagrams,
\begin{equation}
  \pi \circ \sigma 
  = \birdtrack{14ex}{S3__132__12_} 
  = \birdtrack{8ex}{S3__23_} \, , 
\end{equation}
and twist lines at will -- it only matters where lines enter and leave. 
\end{itemize}
Finally, we obtain
\begin{equation}
  \pi \circ \sigma 
  = \begin{pmatrix} 1 & 2 & 3 \\ 1 & 3 & 2 \end{pmatrix}
  = (23) 
  = \birdtrack{8ex}{S3__23_} \ , 
\end{equation}

\vspace{2ex}\noindent{\bf Exercises:} 
\begin{exc}
\item Determine $\sigma\circ\pi$ in three different ways. 
\item Write 
\begin{equation}
  \begin{pmatrix} 1 & 2 & 3 & 4 & 5 \\
                  2 & 4 & 5 & 1 & 3 \end{pmatrix} \in S_5
\end{equation}
in cycle and birdtrack notation. 
\item Digression: Speaking about cycle notation, watch the video
  {\it An Impossible Bet} by {\tt minutephysics},
  \url{www.youtube.com/watch?v=eivGlBKlK6M}, (but not the
  solution!) and come up with a good strategy.
\end{exc}
\vspace{2ex}

Viewing the individual diagrams on the r.h.s. of
Eqs.~\eqref{eq:epseps_free}, \eqref{eq:epseps_free_prop_A3} and
\eqref{eq:A3} as permutations, the total expression is not an element
of the group $S_3$ but of the group algebra $\mathcal{A}(S_3)$. Recall
that the group algebra $\mathcal{A}(G)$ of a finite group $G$ is the
vector space spanned by formal linear combinations of the group
elements, with a multiplication induced from the group multiplication.

We define symmetrisers $S$ and anti-symmetrisers $A$ by 
\begin{equation}
\label{eq:sym_asym}
  S = \frac{1}{n!} \sum_{\pi\in S_n} \pi 
  \qquad \text{and} \qquad
  A = \frac{1}{n!} \sum_{\pi\in S_n} \sign(\pi) \, \pi \, , 
\end{equation}  
and denote them by open and solid bars, respectively, 
\begin{equation}
\label{eq:sym_asym_birdtrack}
   S = \birdtrack{12ex}{Sn_S} 
   \qquad \text{and} \qquad
   A = \birdtrack{12ex}{Sn_A} \ .
\end{equation}
For instance, see also Eq.~\eqref{eq:A3}, 
\begin{equation}
\label{eq:S2andA3}
\begin{split}
  \birdtrack{11ex}{S2_S}
  &= \frac{1}{2} \left( \birdtrack{7ex}{S2_id} 
                        + \birdtrack{8ex}{S2__12_} \right)  \\
  \birdtrack{11ex}{A3}
  &= \frac{1}{3!} \left( \birdtrack{7ex}{S3_id} - \birdtrack{8ex}{S3__12_} 
     - \birdtrack{8ex}{S3__23_} - \birdtrack{8ex}{S3__13_} 
     + \birdtrack{8ex}{S3__123_} + \birdtrack{8ex}{S3__132_} \right) \, .
\end{split}
\end{equation}
Notice that in birdtrack notation the sign of a permutation, $(-1)^K$,
is determined by the number $K$ of line crossings; if more than two
lines cross in a point, one should slightly perturb the diagram before
counting, e.g.
$\birdtrack{4ex}{3lines_crossing} \leadsto
\birdtrack{4ex}{3lines_crossing_pert}$ $(K{=}3)$.

\vspace{2ex}\noindent{\bf Exercise:} 
\begin{exc}
\item Expand $\birdtrack{6ex}{S2_A}$ and $\birdtrack{6ex}{S3_S}$ as in
  Eq.~\eqref{eq:S2andA3}.
\end{exc}
\vspace{2ex}

We use the corresponding notation for partial (anti-)symmetrisation over a
subset of lines, e.g.
\begin{equation}
\begin{split}
\label{eq:partial_AS}
  \raisebox{.5ex}{\birdtrack{11ex}{S3_S12}}
  &= \frac{1}{2} \left( \birdtrack{7ex}{S3_id} 
                        + \birdtrack{8ex}{S3__12_} \right)  
  \qquad \text{or} \\[2ex]
  \raisebox{.5ex}{\birdtrack{14ex}{S3_A13}}
  &= \frac{1}{2} \left( \birdtrack{14ex}{S3_A13_i} - \birdtrack{18ex}{S3_A13_ii} \right)
  = \frac{1}{2} \left( \birdtrack{7ex}{S3_id} - \birdtrack{8ex}{S3__13_} \right) \, .
\end{split}
\end{equation}
The prefactor $1/n!=1/|S_n|$ in Eq.~\eqref{eq:sym_asym} is chosen such that $S^2=S$ and $A^2=A$. 

\vspace{2ex}\noindent{\bf Exercise:} 
\begin{exc}
\item Convince yourself that 
\begin{equation}
  \left( \birdtrack{12ex}{Sn_S} \right)^2 = 
  \birdtrack{16.5ex}{Sn_S_2} = \birdtrack{12ex}{Sn_S}
  \qquad \text{and} \qquad A^2 = A \, .
\end{equation}
\end{exc}
\vspace{2ex}

It follows directly from the definition of $S$ and $A$ that when
intertwining any two lines $S$ remains invariant and $A$ changes by a
factor of $(-1)$, i.e.\
\begin{equation}
  \birdtrack{14ex}{Sn_S_twist} = \birdtrack{12ex}{Sn_S}
  \qquad  \text{and} \qquad 
  \birdtrack{14ex}{Sn_A_twist} = - \birdtrack{12ex}{Sn_A} \, .
\end{equation}
This immediately implies that whenever two (or more) lines connect a
symmetriser to an anti-symmetrizer the whole expression vanishes, e.g.\
\begin{equation}
  \birdtrack{16ex}{SxA} = 0 \, .
\end{equation}
Symmetrisers and anti-symmetrisers can by built recursively. To this
end notice that on r.h.s.\ of
\begin{equation}
  \birdtrack{12ex}{Sn_S_flip} 
  = \frac{1}{n} \left( 
  \raisebox{.7ex}{\birdtrack{12ex}{Sn_Sn-1}} 
  + \raisebox{.7ex}{\birdtrack{14ex}{Sn_Sn-1_twist}} 
  + \ \ldots \  
  + \raisebox{.7ex}{\birdtrack{14ex}{Sn_Sn-1_ttwist}} \right)
\end{equation}
we have sorted the terms according to where the last line is mapped --
to the $n$th, to the $(n{-}1)$th, \ldots, to the first line
line. Multiplying with $\birdtrack{6ex}{Sn_Sn-1}$ from the left and
disentangling lines we obtain the compact relation
\begin{equation}
\label{eq:S_recursive} 
  \birdtrack{12ex}{Sn_S_flip} 
  = \frac{1}{n} \left( 
  \raisebox{.7ex}{\birdtrack{12ex}{Sn_Sn-1}} 
  + (n-1) \raisebox{.7ex}{\birdtrack{19ex}{Sn_SxS}} \right) \, .
\end{equation}
Similarly for anti-symmetrisers: 
\begin{equation}
\label{eq:A_recursive} 
\begin{split}
  \birdtrack{12ex}{Sn_A_flip} 
  &= \frac{1}{n} \left( 
  \raisebox{.7ex}{\birdtrack{12ex}{Sn_An-1}} 
  - \raisebox{.7ex}{\birdtrack{14ex}{Sn_An-1_twist}} 
  + \ \ldots \  
  + (-1)^{n-1} \raisebox{.7ex}{\birdtrack{14ex}{Sn_An-1_ttwist}} \right)
  \\[2ex]
  \birdtrack{12ex}{Sn_A_flip} 
  &= \frac{1}{n} \left( 
  \raisebox{.7ex}{\birdtrack{12ex}{Sn_An-1}} 
  - (n-1) \raisebox{.7ex}{\birdtrack{19ex}{Sn_AxA}} \right) \, .
\end{split}
\end{equation}

\vspace{2ex}\noindent{\bf Exercise:} 
\begin{exc}
\item Convince yourself that the signs in Eq.~\eqref{eq:A_recursive}
  are correct.
\end{exc}
\vspace{2ex}

\subsection{Recap: group algebra and regular representation}

The group algebra $\mathcal{A}(G)$ of a finite group $G$ (i.e.\ the
$\C$-vector space spanned by formal linear combinations of the group
elements, with multiplication induced from the group multiplication),
carries the so-called regular representation of $G$. The regular
representation can be completely reduced to a direct sum containing
all irreducible representations of $G$. Irreducible invariant
subspaces are obtained by right-multiplication with primitive
idempotents $e_j \in \mathcal{A}(G)$. For $G=S_n$ we already know two
such idempotents, the symmetriser $S$ and the anti-symmetriser $A$,
see Eqs.~\eqref{eq:sym_asym} and
\eqref{eq:sym_asym_birdtrack}. Primitive idempotents generating all
irreducible representations of $S_n$ are the so-called Young
operators.

\subsection{Recap: Young operators}
\label{sec:young}

Young diagrams are arrangements of $n$ boxes in $r$ rows of 
non-increasing lengths. A Young tableau $\YTab$ is a Young diagram
with each of the numbers $1,\hdots,n$ written into one of its
boxes. For a so-called standard Young tableau the numbers increase
within each row from left to right and within each column from top to
bottom. We denote the set of all standard Young tableaux with $n$
boxes by $\SYTx_n$, e.g.
\begin{equation}
  \SYTx_2 = \left\{ \scyoung{12} \, , \ \scyoung{1,2} \right\} \, , \quad  
  \SYTx_3 = \left\{ \scyoung{123} \, , \ \scyoung{12,3} \, , \ 
                    \scyoung{13,2} \, , \ \scyoung{1,2,3} \right\} \, .
\end{equation}
Removing the box containing the number $n$ from $\YTab \in \SYTx_n$ we
obtain a standard tableau $\YTab' \in \SYTx_{n-1}$.

For $\YTab\in\SYTx_n$ let $\{h_\YTab\}$ be the set of all horizontal
permutations, i.e. $h_\YTab \in S_n$ leaves the sets of numbers
appearing in the same row of $\YTab$ invariant. Analogously, vertical
permutations $v_\YTab$ leave the sets of numbers appearing in the same
column of $\YTab$ invariant. The Young operator $\YOp_\YTab$ is then
defined in terms of the row symmetrizer,
$s_\YTab = \sum_{\{h_\YTab\}} h_\YTab$, and the column
anti-symmetrizer,
$a_\YTab = \sum_{\{v_\YTab\}} \mathrm{sign}(v_\YTab) v_\YTab$, as
\begin{equation}
\label{eq:definition_Y}
  \YOp_\YTab = \tfrac{1}{|\YTab|} s_\YTab a_\YTab \, .
\end{equation}
Note that as opposed to Eq.~\eqref{eq:sym_asym} we have not included
normalising factorials. The normalisation factor is given by the
product of hook lengths of the boxes of $\YTab$, and thus depends only
the shape of the Young tableau, i.e.\ on the Young diagram. The hook
length of a given box counts the number of boxes below and to the
right of this box, adding one for the box itself. For illustration we
write the hook lengths into the boxes of a couple of Young diagrams,
\begin{equation}
\scyoung{21} \, , \quad 
\scyoung{31,1} \, , \quad 
\scyoung{3,2,1} \, , \quad 
\scyoung{32,21} \, , \quad 
\scyoung{431,21} \, , 
\end{equation}
and calculate the corresponding normalisation factors, 
\begin{equation}
\Yboxdim{1ex}
\left| \yng(2) \right| = 2 \, , \quad 
\left| \yng(2,1) \right| = 3 \, , \quad 
\left| \yng(1,1,1) \right| = 6 \, , \quad 
\left| \yng(2,2) \right| = 12 \, , \quad 
\left| \yng(3,2) \right| = 24 \, .
\end{equation}
Young operators $\YOp_\YTab \in \alg(S_n)$ corresponding to standard
Young tableaux are primitive idempotents. For
$\YTab,\vYTab \in \SYTx_n$ they satisfy
$\YOp_\YTab \YOp_{\vYTab} = 0$, i.e.\ they are transversal, if the
corresponding Young diagrams have different shapes. Unfortunately, for
different Young tableaux of the same shape it can happen that
$\YOp_\YTab \YOp_{\vYTab} \neq 0$ when $n>4$.

In birdtrack notation we can draw Young operators, using partial
(anti-)symmetrisers as introduced in Eq.~\eqref{eq:partial_AS}, e.g.
\begin{equation}
\begin{split}
  \YOp_{_\tyoung{123}} &= \birdtrack{10ex}{S3} \, , \hspace{13ex}
  \YOp_{_\tyoung{1,2,3}} = \birdtrack{10ex}{A3} \, , \\[2ex]
  \YOp_{_\tyoung{12,3}} 
  &= \frac{4}{3} \raisebox{.7ex}{\birdtrack{16ex}{young12-3}} \, , \qquad 
  \YOp_{_\tyoung{13,2}} 
  = \frac{4}{3} \raisebox{.7ex}{\birdtrack{16ex}{young13-2}} \, .
\end{split}
\end{equation}
Note that the normalisation factors are in agreement with
Eq.~\eqref{eq:definition_Y} and the normalisation~\eqref{eq:sym_asym}
of (anti-)symmetrisers. The following $5$-box examples illustrate the
loss of transversality for $n>4$,
\begin{equation} 
\label{eq:5-box-examples}
  \YOp_{_\tyoung{123,45}} 
  = 2 \birdtrack{16ex}{young123-45} \ , 
  \qquad \text{and} \qquad  
  \YOp_{_\tyoung{135,24}} 
  = 2 \birdtrack{16ex}{young135-24} \ , 
\end{equation} 
as we have 
\begin{equation}
\label{eq:transversilty_lost}
  \YOp_{_\tyoung{135,24}} \YOp_{_\tyoung{123,45}} = 0
  \qquad \text{but} \qquad
  \YOp_{_\tyoung{123,45}} \YOp_{_\tyoung{135,24}} \neq 0 \, .
\end{equation} 

\vspace{2ex}\noindent{\bf Exercise:} 
\begin{exc}
\item Verify Eq.~\eqref{eq:transversilty_lost}.
\end{exc}
\vspace{2ex}

\subsection[Young operators and $\SU(N)$: multiplets]{Young operators and $\boldsymbol{\SU(N)}$: multiplets}

Recall that $V\cong\C^{N}$ (or $\overline{V}$) carries the defining
(or complex conjugate) representation of $\SU(N)$. A tensor product
$V^{\otimes n}$ (or $\overline{V}^{\otimes n}$) carries a product
representation of $\SU(N)$. This tensor product also carries a
representation of $S_n$. The representations of these two groups
commute, since $\SU(N)$ acts only on individual factors, on each in
the same way, whereas $S_n$ acts by permuting the factors. In fact, it
is a standard result, that Young operators viewed as linear maps
$V^{\otimes n} \to V^{\otimes n}$ (or
$\overline{V}^{\otimes n} \to \overline{V}^{\otimes n}$) project onto
irreducible $\SU(N)$-invariant subspaces.

In birdtrack notation this means that we simply add arrows to the
lines in all diagrams for Young operators, all pointing in the same
direction, e.g.,
$\YOp_{_\tyoung{12,3}}: V^{\otimes n} \to V^{\otimes n}$ reads
\begin{equation}
\label{eq:Young_with_arrows}
  \YOp_{_\tyoung{12,3}} 
  = \frac{4}{3} \raisebox{.7ex}{\birdtrack{16ex}{young12-3_sworra}} \, , 
\end{equation}
and since all arrows point in the same direction, we usually
immediately drop them again. 

We refer to irreducible $\SU(N)$-invariant subspaces as multiplets,
e.g., a one-dimensional subspace (carrying the trivial representation)
is called singlet. The dimension of a multiplet is given by the trace
of the projector, e.g.
\begin{equation}
\begin{split}
 \utr \YOp_{_\tyoung{12}} 
 &= \raisebox{-.7ex}{\birdtrack{15ex}{tr_young_12}} 
 = \frac{1}{2} \left( \birdtrack{15ex}{tr_young_12_01} 
   + \birdtrack{15ex}{tr_young_12_02} \right)
 = \frac{1}{2}(N^2+N) \\[2ex]
 &= \frac{N(N+1)}{2} \, ,
\end{split}
\end{equation}
or
\begin{equation}
\begin{split}
 \utr \YOp_{_\tyoung{12,3}} 
 &= \frac{4}{3} \birdtrack{15ex}{tr_young12-3} 
 = \frac{2}{3} \left( \birdtrack{15ex}{tr_young12-3_01} 
   - \birdtrack{15ex}{tr_young12-3_02_NEW} \right)\\[2ex]
 &= \frac{1}{3}\left(N^2(N+1)-N(N+1)\right) 
 = \frac{N}{3}(N^2-1) \, ,
\end{split}
\end{equation}
For $N=3$ the latter describes an octet, which actually carries the
adjoint representation of $\SU(3)$.

\vspace{2ex}\noindent{\bf Exercise:} 
\begin{exc}
\item The Young diagram for the adjoint representation of $\SU(N)$ is
  given by a column of $N{-}1$ boxes and a column with one box -- why?
  Verify that the dimension of the adjoint representation is $N^2{-}1$
  by calculating $\utr \YOp_{_\tyoung{1N,2,\dirtyvdots}}$ in
  birdtracks.\\
  Hint: First expand the symmetriser, then evaluate the trace of the
  remaining anti-symmetriser using the recursion
  relation~\eqref{eq:A_recursive}.
\end{exc}
\vspace{2ex}


\section{Colour space}
\label{sec:color_space}

Consider a QCD process with a couple of incoming and outgoing quarks,
anti-quarks, and gluons. A corresponding Feynman diagram might look
like this,
\begin{equation}
\birdtrack{42ex}{color_structure} \, , 
\end{equation}
where inside some of the lines are connected directly, whereas others
are tied together according to the QCD Feynman rules, i.e.\ by
quark-gluon vertices as well as by 3-gluon and 4-gluon vertices. Note
that these rules ensure that there are always as many
(anti\mbox{-})quark lines with arrows pointing away from the central
blob as there are lines with arrows pointing towards it. The amplitude
corresponding to such a diagram is the product of a kinematic factor
(some often initially divergent momentum space integral), a spin
factor and a colour factor. Here, we are only interested in the colour
factor, the so-called colour structure.

The colour structure is a tensor
$c \in (\overline{V} \otimes V)^{\otimes n_q} \otimes A^{\otimes
  n_g}$.
Since colour is confined the only relevant colour structures are
singlets, i.e.\ tensors $c$ which transform in the trivial
representation of $\SU(N)$. In other words, $c$ is a so-called
invariant tensor. The singlet-subspace of
$(\overline{V} \otimes V)^{\otimes n_q} \otimes A^{\otimes n_g}$ is
called colour space.

Quark lines $ $, gluon lines $ $ and quark-gluon vertices (generators)
$ $ are examples of invariant tensors. Any combination (tensor
product, contraction) of invariant tensors is also an invariant
tensor. Thus, Young operators and triple gluon vertices (structure
constants),
\begin{equation}
\label{eq:f_as_loop_A}
  \birdtrack{12ex}{structure_constants_no_indices} 
  = \frac{2}{T_R} \birdtrack{28ex}{f_as_loop_A} \, , 
\end{equation}
cf.~Eq.~\eqref{eq:f_as_3loops}, are invariant tensors, and also the
fully symmetric
\begin{equation}
  \birdtrack{12ex}{d_vertex} 
  := \frac{2}{T_R} \birdtrack{28ex}{d_as_loop_S} \, .
\end{equation}

The vector space $V \cong \C^N$, which carries the defining representation
of $\SU(N)$, is endowed with a scalar product, which is also left
invariant by $\SU(N)$. This scalar product also induces scalar
products on $\overline{V}$, $A$ and on arbitrary tensor products of
$V,\overline{V}$ and $A$. For two colour structures
$c_1,c_2 \in (\overline{V} \otimes V)^{\otimes n_q} \otimes A^{\otimes n_g}$
we have
\begin{equation}
\label{eq:sp_color_space}
 \langle c_1 , c_2 \rangle = \utr (c_1^\dag c_2) 
 = \left(((c_1)^{ab\ldots})\indices{^{jk\ldots}_{\ell m\ldots}}\right)^*
   ((c_2)^{ab\ldots})\indices{^{jk\ldots}_{\ell m\ldots}} \, ,
\end{equation}
where in birdtrack notation Hermitian conjugation corresponds to
mirroring the diagram across a vertical line and reversing all arrows.

For calculations it is convenient to expand colour structures into a
basis of colour space. The most popular bases in use are so-called
trace bases (which in general are overcomplete, i.e.\ they are no
proper bases but only spanning sets). In general, trace bases are not
orthogonal. In the remainder of this section, after a brief review of
trace bases, we will discuss the construction of minimal, orthogonal
bases, so-called multiplet bases.

\subsection{Trace bases vs. multiplet bases}
\label{subsec:trace_vs_multiplet}

{\bf Trace bases.} For diagrams with a given number of external
(anti\mbox{-})quark lines and gluon lines, the trace basis can be
constructed as follows: 
\begin{itemize}[itemsep=1ex,topsep=1ex,parsep=0mm,partopsep=0mm]
\item Attach a quark-gluon vertex to each external gluon line. 
\item Then connect all external (anti\mbox{-})quark lines to either
  external(anti\mbox{-})quark lines or (anti\mbox{-})quark lines from
  quark-gluon vertices.
\end{itemize}
Examples:\\[.5ex]
For $\overline{V} \otimes V \otimes A^{\otimes 2}$ we have to connect
the (anti\mbox{-})quark lines ending on the dashed box,
\begin{equation}
  \birdtrack{26ex}{VbarVA_2_tracebasis} \, ,
\end{equation}
in all possible ways. The non-zero possibilities are 
\begin{equation}
  \birdtrack{26ex}{VbarVA_2_tracebasis_connect_1} \, , \quad 
  \birdtrack{26ex}{VbarVA_2_tracebasis_connect_2} \, , \quad 
  \birdtrack{26ex}{VbarVA_2_tracebasis_connect_3} \, , 
\end{equation}
yielding
\begin{equation}
\label{eq:trace_basis_VbarVA^2}
  c_1 = \birdtrack{20ex}{VbarVA_2_tracebasis_1} \, , \quad 
  c_2 = \birdtrack{14ex}{VbarVA_2_tracebasis_2} \, , \quad 
  c_3 = \birdtrack{18ex}{VbarVA_2_tracebasis_3} \, , 
\end{equation}
where we have omitted irrelevant prefactors. Note that in general the
$c_j$ are not mutually orthogonal, e.g.\
\begin{equation}
  \langle c_1,c_2 \rangle 
  = \utr(c_1^\dag c_2) 
  = \birdtrack{8ex}{trCF} 
  = T_R (N^2-1) 
  = C_F N \, , 
\end{equation}
see Eqs.~\eqref{eq:remove_gluon} and \eqref{eq:C_F}. 
\\[.5ex]
For $A^{\otimes 4}$, connecting the (anti\mbox{-})quark lines
ending on the dashed box,
\begin{equation}
  \birdtrack{26ex}{A_4_tracebasis} \, , 
\end{equation}
in all possible ways, we in turn find the non-vanishing diagrams
\begin{equation}
\label{eq:A^4_trace_basis}
\begin{split}
  &\birdtrack{12ex}{A_4_id} \, , \quad 
  \birdtrack{12ex}{A_4_cross} \, , \quad 
  \birdtrack{12ex}{A_4_tr} \, , \quad\\ 
  &\birdtrack{18ex}{A_4_loop} \, , \quad 
  \birdtrack{18ex}{A_4_pool} \, , \quad 
  \birdtrack{16ex}{A_4_eight} \, , \quad 
  \birdtrack{16ex}{A_4_thgie} \, , \quad\\
  &\birdtrack{28ex}{A_4_inf} \, , \quad 
  \birdtrack{28ex}{A_4_fni} \, , 
\end{split}
\end{equation}
again omitting all prefactors. Written in index (or matrix) notation
these basis vectors consist of traces of products of generators, thus
the name trace basis. The colour structures~\eqref{eq:A^4_trace_basis}
are again not mutually orthogonal, e.g.
\begin{equation}
  \left\langle \birdtrack{8ex}{A_4_id} \, , \, 
               \birdtrack{8ex}{A_4_tr} \right\rangle
  = \birdtrack{6ex}{gluon_loop}
  = N^2-1 \, .
\end{equation}
Moreover, for $N=3$ the nine
$A^{\otimes 4}$-vectors~\eqref{eq:A^4_trace_basis} are not linearly
independent, since, as we will see below, the colour space which they
span is only eight-dimensional.

There is a simple algorithm, for writing arbitrary colour factors as
linear combinations of trace basis elements: First replace all
four-gluon vertices by (one gluon-contracted linear combinations of)
three-gluon vertices. Then replace all three-gluon vertices by (linear
combinations of) quark loops with three gluons attached, see
Eqs.~\eqref{eq:f_as_3loops} or \eqref{eq:f_as_loop_A}. Finally, remove
all internal gluon lines using Eq.~\eqref{eq:remove_gluon}.

\vspace{2ex}\noindent
{\bf Multiplet bases.} For the construction of a multiplet basis for
$c \in (\overline{V} \otimes V)^{\otimes n_q} \otimes A^{\otimes
  n_g}$ we consider $c$ as a linear map, say
\begin{equation}
\label{eq:c_as_linear_map}
  c: (\overline{V} \otimes V)^{\otimes k_q} \otimes A^{\otimes
  k_g} \to (\overline{V} \otimes V)^{\otimes (n_q-k_q)} \otimes A^{\otimes
  (n_g-k_g)} \, ,
\end{equation}
for some $0\leq k_q \leq n_q$ and $0\leq k_g \leq n_g$. 
In general, we thus have a linear map $c: W_1 \to W_2$, between two
vector spaces, carrying representations $\Gamma_1$ and $\Gamma_2$ of
$\SU(N)$. Moreover, $c$ being an invariant tensor means that
\begin{equation}
  c \circ \Gamma_1(g) = \Gamma_2(g) \circ c \quad \forall\, g \in \SU(N) \, .
\end{equation}
Now we are in a situation where we can employ Schur's lemma. It is
often formulated for the case where $W_1$ and $W_2$ carry irreducible
representations saying that 
\begin{itemize}[itemsep=1ex,topsep=1ex,parsep=0mm,partopsep=0mm]
\item if the two representations are inequivalent, then $c$ vanishes
  identically, and 
\item if the two representations are equivalent and $W_1=W_2$, then
  $c$ is a multiple of the identity.
\end{itemize}
In our case the representations are typically not irreducible, and
then Schur's lemma implies that $c$ can only map subspaces onto each
other that carry the same irreducible representation, i.e.\ $c$ maps
only equivalent multiplets onto each other.

Now consider the case when $W_1=W_2=:W$. 
If we decompose $W$ into multiplets, i.e.\ into irreducible
$\SU(N)$-invariant subspaces, then the projectors onto these
multiplets are distinguished elements of colour space.
\begin{itemize}[itemsep=1ex,topsep=1ex,parsep=0mm,partopsep=0mm]
\item If each multiplet appears only once in the decomposition of $W$
  then the projectors form a basis of colour base. If, moreover, the
  projectors are Hermitian, then this basis is orthogonal. 
\item If some multiplets in the decomposition of $W$ have a
  multiplicity $>1$ then we have to complement the projectors with
  operators mapping equivalent multiplets onto each other.
\end{itemize}
In practice, finding the multiplets in the decomposition of $W$ and
their multiplicities can, e.g., be done by multiplying Young diagrams
according to the standard rules. The crucial step is then to find
Hermitian projectors onto these multiplets. Finally, multiplet bases
can be constructed straightforwardly from Hermitian projection
operators.

In the following sections we discuss how to construct Hermitian
projection operators as well as multiplet bases for the cases
$V^{\otimes n} \to V^{\otimes n}$ and
$A^{\otimes n} \to A^{\otimes n}$. Moreover, we will see that
multiplet cases for any colour space can be constructed from
projectors for $A^{\otimes n} \to A^{\otimes n}$.

\vspace{2ex}\noindent 
{\bf Comparison.} Trace bases are convenient since they are easy to
construct and since there is a simple algorithm for expanding
arbitrary colour factors into a trace basis. In general, trace bases
are overcomplete, i.e.\ expansions tend to have too many terms. For
instance, the trace basis for the colour space within $A^{\otimes n}$
is a proper basis if $n\leq N$ but for $n>N$ it is only a spanning set
-- the basis vectors are linearly dependent. Trace bases, typically,
are also non-orthogonal.

Constructing multiplet bases requires more work than constructing
trace bases. In return we obtain not only a proper basis, i.e.\ the
basis vectors are linearly independent, but also an orthogonal
basis. Even though the dimension of colour space depends on $N$, the
number of colours, the birdtrack construction of multiplet bases can
be carried out independently of $N$, and then for small $N$ some basis
vectors simply vanish.

The numbers in Table~\ref{tab:some_numbers} give us an impression of
the potential advantage of multiplet bases over trace bases. Imagine
doing a calculation for $N=3$ with $6$ to $10$ external gluons
involved. Then the number of trace basis elements exceeds the
dimension of colour space by, roughly, a factor of $2$ to
$12$. Expanding colour structures in a trace or multiplet basis and
then, e.g., calculating scalar products will result in $4$ to $144$
times as many terms when using a trace basis instead of a multiplet
basis.

\begin{table}[t]
\begin{center}\small
\begin{tabular}{|l c c c c|} 
\hline
\hspace*{15ex}& \multicolumn{2}{c}{\hspace*{-.5ex}Number of multiplets\hspace*{4ex}} & \multicolumn{2}{c|}{\hspace*{-2ex}Dimension of colour space}\\
& $N=3$ & $N=\infty$ & $N=3$ & $N=\infty$ \\[0.5ex] 
\hline &&&& \\[-2ex]
$A^{\otimes 2} \to A^{\otimes 2}$ & 6 & 7 & 8 & 9 \\ 
$A^{\otimes 3} \to A^{\otimes 3}$ & 29 & 51 & 145 & 265  \\
$A^{\otimes 4} \to A^{\otimes 4}$ & 166 & 513 & 3 598 & 14 833\\
$A^{\otimes 5} \to A^{\otimes 5}$ & 1 002 & 6 345 & 107 160 & 1 334 961\\[.3ex] 
\hline
\end{tabular}
\caption{Number of projection operators and dimension of the colour
  space within $A^{\otimes(2n)}$, for colour structures viewed as maps
  $A^{\otimes n} \to A^{\otimes n}$. The first two columns show the
  number of multiplets (counted with multiplicities) in the
  decomposition of $A^{\otimes n}$, both for $N=3$ and for $N\geq n$.
  The last two columns contain the dimensions of the respective colour
  spaces; the dimension in the last column is also equal to the number of
  elements of the corresponding trace basis.}
\label{tab:some_numbers}
\end{center}
\vspace*{-4ex}
\end{table}

\subsection{Multiplet bases for quarks}

We first consider the case without external gluons, i.e.\ we are
interested in the colour space within $(\overline{V}\otimes V)^{n}$.
Tensors $c \in (\overline{V}\otimes V)^{\otimes n}$ can be viewed as
linear maps $c: V^{\otimes n} \to V^{\otimes n}$, and Young operators
$\YOp_\YTab$ project onto multiplets. Unfortunately, Young operators
are in general not Hermitian, as can be seen by, e.g.\, inspecting
Eq.~\eqref{eq:Young_with_arrows}: Mirroring and reversing the arrows
does not yield back the original expression.

However, Hermitian operators $\HYOp_\YTab$ corresponding to standard
Young tableaux $\YTab$ can be constructed. In \cite{Cvi08} they are
derived as solutions of certain characteristic equations. They can
also be written down directly starting from a Young tableaux as
follows. Consider the sequence of Young tableaux $\YTab_j \in \SYTx_j$
obtained from $\YTab\in\SYTx_n$ by, step by step, removing the box
with the highest number, e.g., starting with
$\YTab=\YTab_3=\tyoung{12,3}\in\SYTx_3$ we obtain
\begin{equation}
\label{eq:Young_construction_history}
  \YTab_1=\scyoung{1} \, , \quad 
  \YTab_2=\scyoung{12} \, , \quad 
  \YTab_3=\scyoung{12,3} \, .
\end{equation}
Young operators for $n=2$ are Hermitian -- they are just total
(anti-)symmetrisers -- so we set \
\begin{equation}
  \HYOp_{\YTab_j} = \YOp_{\YTab_j} \quad \forall\, j \leq 2\, .
\end{equation}
Then we define recursively
\begin{equation}
\label{eq:Hermitian_Young_recursion}
  \HYOp_{\YTab_j} 
  = (\HYOp_{\YTab_{j-1}} \otimes \eins_V)
    \YOp_{\YTab_j} (\HYOp_{\YTab_{j-1}} \otimes \eins_V)
  \quad \forall\, j \geq 3 \, , 
\end{equation}
i.e.\ in birdtrack notation we take the Young operator
$\YOp_{\YTab_j}$ and write the Hermitian Young operator
$\HYOp_{\YTab_{j-1}}$ over the first $j{-}1$ lines, to the left and to
the right. For instance,
\begin{equation}
  \HYOp_{_\tyoung{12,3}} 
  = \frac{4}{3} \birdtrack{27ex}{young12-3_sandwiched} 
  = \frac{4}{3} \birdtrack{22ex}{Hyoung12-3} \, , 
\end{equation}
which is manifestly Hermitian, and in birdtracks it is also easy to
see that
\begin{equation}
  \utr \HYOp_{_\tyoung{12,3}} = \utr \YOp_{_\tyoung{12,3}}
\end{equation}
since $\left( \birdtrack{5ex}{S2_S}\right)^2=\birdtrack{5ex}{S2_S}$. 

It can be shown \cite{KepSjo14} that the resulting $\HYOp_\YTab$ not
only project onto the correct multiplets but that they are also
Hermitian and thus mutually orthogonal with respect to the scalar
product~\eqref{eq:sp_color_space}. Furthermore, using the Hermitian
Young operators $\HYOp_\YTab$ automatically cures the loss of
transversality mentioned at the end of Sec.~\ref{sec:young}

The recursive construction can produce initially lengthy expressions
which can often be simplified considerably, see, e.g., the step-by-step
example for $\tyoung{135,24}$ in the Appendix of \cite{KepSjo14}.
Similar simplifications can be shown to occur much more generally
\cite{AlcWei17a} and they can be used to devise a recipe for directly
writing down fully simplified Hermitian Young operators
\cite{AlcWei17b}.

With the Hermitian Young operators 
\begin{equation}
  \HYOp_{_\tyoung{123}} = \birdtrack{7ex}{S3} \, , \quad 
  \HYOp_{_\tyoung{12,3}} = \frac{4}{3} \birdtrack{15ex}{Hyoung12-3} \, , \quad 
  \HYOp_{_\tyoung{13,2}} = \frac{4}{3} \birdtrack{15ex}{Hyoung13-2} \, , \quad 
  \HYOp_{_\tyoung{1,2,3}} = \birdtrack{7ex}{A3} \, , 
\end{equation}
we have completely decomposed $V^{\otimes3}$ into an orthogonal sum of
multiplets. However, the $\HYOp_\YTab$ alone do not form a basis for
the colour space within $(\overline{V}\otimes V)^{\otimes3}$, since the
multiplet $\Yboxdim{1ex}\yng(2,1)$ appears twice, i.e.\ we also need
an operator mapping these two multiplets onto each other. To this
end we write down $\HYOp_{_\tyoung{12,3}}$ and $\HYOp_{_\tyoung{13,2}}$
next to each other (omitting prefactors),
\begin{equation}
  \birdtrack{45ex}{12-3_dashed_13-2}
  \, , 
\end{equation}
and seek a way of connecting the lines within the dashed box such that
the whole expression does not vanish, because then it is guaranteed,
that the resulting expression has the same kernel as
$\HYOp_{_\tyoung{13,2}}$ and the same image as
$\HYOp_{_\tyoung{12,3}}$. The only such connection (up to a sign) is 
\begin{equation}
\label{eq:no_zero_3q_transition}
  \birdtrack{40ex}{12-3_connected_13-2} \, , 
\end{equation}
and by expanding the central (anti-)symmetrisers one can verify that
this expression is proportional to
\begin{equation}
\label{eq:3q_transition_basis_vector}
  T_1 := \birdtrack{16ex}{12-3_to_13-2} \, . 
\end{equation}
Thus, we have found a basis vector mapping $\tyoung{13,2}$ to
$\tyoung{12,3}$. The vector for the reverse mapping can be obtained in
the same way and reads
\begin{equation}
  T_2 := \birdtrack{16ex}{13-2_to_12-3} \, . 
\end{equation}

\vspace{2ex}\noindent{\bf Exercise:} 
\begin{exc}
\item Define 
\begin{equation}
  B =  \birdtrack{16ex}{12-3_to_13-2_x__23_} \, , 
\end{equation}
and show that $B^2$ is proportional to $B$ by expanding the central
(anti-)symmetrisers. Explain why this implies that the birdtrack
diagram~\eqref{eq:no_zero_3q_transition} is proportional to $T_1$.
\end{exc}
\vspace{2ex}

The multiplet basis for $V^{\otimes3} \to V^{\otimes3}$ consisting of
four Hermitian Young operators and two transition operators is
orthogonal.  If desired the basis vectors can be normalised: For
Hermitian projection operators we generally have
\begin{equation}
  \langle \HYOp_\YTab, \HYOp_\YTab \rangle 
  = \utr (\HYOp_\YTab^\dag \HYOp_\YTab) 
  = \utr (\HYOp_\YTab \HYOp_\YTab) 
  = \utr (\HYOp_\YTab)
  = \dim M_\YTab
\end{equation}
where $M_\YTab$ is the multiplet to which $\HYOp_\YTab$ projects. Hence,
\begin{equation} 
\label{eq:normalise_projector}
  \tfrac{1}{\sqrt{\dim M_\YTab}} \, \HYOp_\YTab
\end{equation}
has norm one. The transition operators can straightforwardly be
normalised by a direct calculation.

\vspace{2ex}\noindent{\bf Exercise:} 
\begin{exc}
\item Calculate $\langle T_1, T_1 \rangle$ and normalise $T_1$
  accordingly.
\end{exc}
\vspace{2ex}

\subsection{General multiplet bases}
\label{subsec:general_bases}

\vspace{2ex}\noindent {\bf Gluons only.} The construction of multiplet
bases for 
an even number of external gluons has been outlined in
Sec.~\ref{subsec:trace_vs_multiplet}: Consider colour structures as
maps $A^{\otimes n} \to A^{\otimes n}$, construct Hermitian projection
operators to multiplets within $A^{\otimes n}$, and complement the
projection operators with transition operators between equivalent
multiplets.

For an odd number of external gluons, i.e.\ when considering the
colour space within $A^{\otimes(2n+1)}$ we consider colour structures
as maps $A^{\otimes n} \to A^{\otimes (n+1)}$. In
Sec.~\ref{subsec:gluon_projectors} we will see that knowing the
projectors to multiplets within $A^{\otimes n}$ allows to
construct the projectors to all equivalent multiplets
within $A^{\otimes (n+1)}$ in a straightforward way.\footnote{Using
  the terminology of Sec.~\ref{subsec:gluon_projectors} only projectors
  to old multiplets have to be constructed.} Transition operators
between equivalent multiplets within $A^{\otimes n}$ and
$A^{\otimes (n+1)}$ then constitute the desired multiplet basis.

\vspace{2ex}\noindent {\bf Quarks and gluons.} When (anti-)quark lines
are present we can always group together an anti-quark line with a
quark line. Consider, e.g., the colour space within
$\overline{V}\otimes V \otimes A^{\otimes n}$. Noting that
$\overline{V} \otimes V = \bullet \oplus A$, see
Eq.~\eqref{eq:decomposeVbarV}, the quark-anti-quark pair can be either
in a singlet state or in a state transforming in the adjoint
representation. In the singlet case we have to study the colour space
within $A^{\otimes n}$, in the adjoint case the colour space within
$A^{\otimes (n+1)}$.

\vspace{2ex}\noindent 
{\bf Conclusion:} General multiplet bases can be constructed in a
straightforward way from gluon projectors. In particular, the gluon
projectors for $A^{\otimes \nu} \to A^{\otimes \nu}$,
$\nu=0,\ldots,n$, are sufficient for constructing multiplet bases for
the colour spaces within
$\big(\overline{V}\otimes V\big)^{\otimes k} \otimes A^{\otimes(2n+1-k)}$ with
arbitrary $k=0,\ldots,2n{+}1$.

\subsection{Gluon projectors}
\label{subsec:gluon_projectors} 

In Sec.~\ref{subsec:general_bases} we have seen that the crucial
ingredient for any multiplet basis are the projection operators to
multiplets within $A^{\otimes n}$.

The construction rules for projectors depend on when a multiplet $M$
appears for the first time in the sequence
\begin{equation}
  A^{\otimes 0} = \bullet \, , \ 
  A^{\otimes 1}=A \, , \
  A^{\otimes 2} = A \otimes A \, , \
  A^{\otimes 3} \, , \
  A^{\otimes 4} \, , \ldots
\end{equation}
We call $n_f(M) = 0,1,2,3,4,\ldots$ the first occurrence of
multiplet~$M$. Consequently, the only multiplets with first occurrence
$0$ and $1$ are the trivial and the adjoint representation,
respectively, in short $n_f(\bullet)=0$ and $n_f(A) = 1$. For
$\SU(3)$, we have, e.g.,
\begin{eqnarray}
\label{eq:SU388}
\Yboxdim{1.5ex}
  \begin{array}{ccccccccccccccc}
    \yng(2,1)   
    & \otimes 
    &\yng(2,1)  
    & \ = \ 
    & \bullet 
    & \oplus 
    & \yng(2,1) 
    & \oplus 
    & \yng(2,1) 
    & \oplus 
    & \yng(3) 
    & \oplus 
    &  \yng(3,3)
    & \oplus 
    &  \yng(4,2)
    \\[1ex]
    8
    & & 8 
    & & 1 
    & & 8\,
    & & 8 \,
    & & 10
    & & \overline{10} 
    & & 27
  \end{array}
  \, ,
\end{eqnarray}
and, consequently, the decuplets and the $27$-plet have first
occurrence $2$. The following table shows some more $\SU(3)$ examples.
\vspace{2ex}
\begin{center}
\begin{tabular}{|c||c|c|c|c|}
\hline
&&&&\\[-2ex]
$n_f$ & $0$ & $1$ & $2$ & $3$\\[.3ex]
\hline&&&&\\[-1.7ex]
$\SU(3)$ & 
$\Yboxdim{1.5ex}\bullet=\yng(1,1,1)$ & 
$\Yboxdim{1.5ex}A=\yng(2,1)$ & 
$\Yboxdim{1.5ex}\yng(3)$ & 
$\Yboxdim{1.5ex}\yng(5,1)$ \\
Young diagrams & & & 
$\Yboxdim{1.5ex}\yng(3,3)$ & 
$\Yboxdim{1.5ex}\yng(5,4)$ \\[2mm]
& & & $\Yboxdim{1.5ex}\yng(4,2)$ & 
$\Yboxdim{1.5ex}\yng(6,3)$\\[2mm]
\hline
\end{tabular}
\end{center}
\vspace{2ex} The first occurrence of any multiplet can in principle be
determined by repeatedly multiplying Young diagrams for the adjoint
representation until the desired multiplet appears. One can also
derive \cite[App.~B]{KepSjo12} a graphical rule for directly determining
$n_f(M)$ from the corresponding Young diagram.

Our construction of gluon projectors will be recursive. Assume that we
have determined the projectors for the decomposition of
$A^{\otimes (n-1)}=\bigoplus_j M_j$ into multiplets $M_j$. In order to
decompose $A^{\otimes n}$ we have to multiply each
$M \subseteq A^{\otimes (n-1)}$ with another $A$, i.e.\ we consider
\begin{equation}
\label{eq:MtimesA}
  M \otimes A  = \bigoplus_k M_k' \, , 
\end{equation}
which for the projectors reads
\begin{equation}
\birdtrack{18ex}{PMxA_NEW} = \sum_k \birdtrack{21ex}{PMprime_NEW} \, .
\end{equation}
As for the Hermitian Young operators, cf.\ the discussion around
Eqs.~\eqref{eq:Young_construction_history}--\eqref{eq:Hermitian_Young_recursion},
our projectors will be such that $\forall\, k$
\begin{equation}
\raisebox{-2ex}{\birdtrack{24ex}{PMprime_par_tr_NEW}}
\quad \text{is proportional to} \qquad\! 
\birdtrack{18ex}{PM_NEW} \, .
\end{equation}
For the decomposition~\eqref{eq:MtimesA} one can show \cite{KepSjo12} that 
\begin{enumerate}[label=(\roman*),itemsep=.5ex,topsep=.5ex,parsep=0mm,
  partopsep=0mm]
\item $n_f(M_k') = n_f(M){-}1$, $n_f(M)$ or $n_f(M){+}1$, and
\item only $M$ itself can appear with multiplicity greater than one
within $M \otimes A$ (in fact it can appear up to $N{-}1$ times), all
other multiplets are unique.
\end{enumerate}
Reexamining Eq.~\eqref{eq:SU388}, where on the l.h.s.\ we identify
$\Yboxdim{1ex} \yng(2,1) \otimes \yng(2,1) = M \otimes A$, we can
verify both statements: Property (i) is trivially true since
$n_f(M)=2$, but we also see that (ii) holds, as on the r.h.s. all
multiplets except for $\Yboxdim{1ex} M=\yng(2,1)$ appear only once, and
$\Yboxdim{1ex} M=\yng(2,1)$ itself appears twice (which here is the
maximum degeneracy since $N=3$).

We call a multiplet $M \subseteq A^{\otimes n}$ old if $n_f(M)<n$ and
new if $n_f(M)=n$. Construction rules for projectors onto multiplets
$M_k' \subseteq M \otimes A$ depend on whether $M$ and $M'$ are old or
new and on which of the cases (i) we have at hand. We now give some
examples for important cases, which all appear in $A \otimes A$; the
complete set of construction rules (and their proofs) are given in
\cite{KepSjo12}.

\vspace{2ex}\noindent 
{\bf $\boldsymbol{M}$ new, 
  $\boldsymbol{n_f(M_k')=n_f(M){-}1}$.}
Write down $P_M$ twice and bend back the last gluon line, 
\begin{equation}
  P_{M_k'} = \frac{\dim M_k'}{\dim M} \birdtrack{48ex}{new_to_lower_NEW} \, ;
\end{equation}
the prefactor makes sure that $P_{M_k'}^2=P_{M_k'}$. Example: The
projector to the singlet in Eq.~\eqref{eq:SU388} is constructed in
this way,
\begin{equation}
  P_\bullet = \frac{1}{N^2-1} \birdtrack{12ex}{A_4_tr} \, .
\end{equation}

\vspace{2ex}\noindent 
{\bf $\boldsymbol{M}$ new, 
$\boldsymbol{M_k'}$ equivalent to $\boldsymbol{M}$.}
Write down $P_M$ twice, connect the new gluon line to one of
the old gluon lines by means of two $f$- or $d$-vertices (or a linear
combination of them), and write another $P_M$ over the central lines,
\begin{equation}
  P_{M_k'} = \,\gamma\, \birdtrack{63ex}{back_to_same_NEW} \, ,
\end{equation}
where \raisebox{-.5ex}{$^\otimes$} is to be replaced by $\bullet$ or
$\circ$ (or a linear combination). A formula for the normalisation
factor $\gamma$ is given in \cite{KepSjo12}. Examples: The two
projectors to copies of the adjoint representation on the r.h.s. of
Eq.~\eqref{eq:SU388} are constructed in this way,
\begin{equation}
  P_{Aa} = \frac{1}{2N} \birdtrack{14ex}{PAa}
  \qquad \text{and} \qquad 
  P_{As} = \frac{1}{2(N^2-4)} \birdtrack{14ex}{PAs} \, .
\end{equation}

\vspace{2ex}\noindent {\bf New multiplets.} Here we use that
$A \subset \overline{V} \otimes V$: Split each gluon line into a quark
and an anti-quark line by means of a generator
(quark-gluon-vertex). Then put a Young operator (Hermitian or not
doesn't matter) on the quark lines and on the anti-quark-lines (one
each),
\begin{equation}
  T = \birdtrack{36ex}{new_multiplet} \, .
\end{equation}
One can show \cite{KepSjo12} that for $\YTab,\YTab'\in\SYTx_n$ the
tensor product $\overline{\YTab} \otimes \YTab'$ contains exactly one
new multiplet of the decomposition of $A^{\otimes n}$. Since it also
contains contributions from other (old) multiplets, these have to be
removed in a Gram-Schmidt step, 
\begin{equation}
  \widetilde{T} 
  = T - \sum_{M\,\text{old}} \frac{\utr(P_M T)}{\dim M} \, P_M \, .
\end{equation}
Finally, we obtain the desired projector onto the new multiplet by
normalising $\widetilde{T}$,
\begin{equation}
  P_{M_k'} = \frac{\dim M_k'}{\utr \widetilde{T}} \, \widetilde{T} \, .
\end{equation}
Examples: Choosing $\Yboxdim{1ex}(\yng(2),\yng(2))$ for the pair
$(\YTab,\YTab')$ this procedure leads to the projector onto the
27-plet in Eq.~\eqref{eq:SU388}. Projectors onto the two decuplets in
Eq.~\eqref{eq:SU388} we find by choosing
$\Yboxdim{1ex}(\yng(2),\yng(1,1))$ and
$\Yboxdim{1ex}(\yng(1,1),\yng(2))$. The resulting formulae are, e.g.,
given in \cite[Table~9.4]{Cvi08}, \cite[App.~A.1]{DokMar06} or
\cite[Eq.~(1.23)]{KepSjo12}.

\vspace{1ex}

Note that all birdtrack construction rules in this section never use
that $N=3$. Thus, the projection operators constructed above, project
onto multiplets within $A\otimes A$ for any~$N$. However, for $N\geq4$
there is exactly one more multiplet in the decomposition of
$A\otimes A$. We find the corresponding projector by applying the
construction rules for new multiplets, this time using
$\Yboxdim{1ex}(\YTab,\YTab')=(\yng(1,1),\yng(1,1))$; the result can be
shown to vanish for $N=3$. This is a general feature of these
birdtrack constructions for gluon projectors (and for multiplet
bases): The construction rules are independent of $N$. If for small
$N$ there are fewer multiplets (or colour spaces of smaller dimension)
then some of the terms simply vanish -- as opposed to less obvious
linear dependencies, which appear in trace bases,
cf.~Eq.~\eqref{eq:A^4_trace_basis}.

\vspace{2ex}\noindent{\bf Exercise:} 
\begin{exc}
\item Construct the projectors $\Yboxdim{1ex}P_{\yng(3)}$,
  $\Yboxdim{1ex}P_{\yng(3,3)}$ and $\Yboxdim{1ex}P_{\yng(4,2)}$
  according to the rules given above. You may either use
  $\Yboxdim{1ex}\overline{P_{\yng(3)}}=P_{\yng(3,3)}\vphantom{f_{f_{f_f}}}$
  in your calculations or verify this property from your result. Also
  determine the dimensions of the corresponding multiplets for
  arbitrary $N$.
\end{exc}

\subsection{Some multiplet bases}

{\bf $\boldsymbol{A^{\otimes4}}$.} The multiplet basis for the colour
space within $A^{\otimes4}$ is given by the projection operators --
six for $N=3$, seven for $N\geq4$ -- normalised according to
Eq.~\eqref{eq:normalise_projector}, and two transition operators
mapping the two multiplets carrying the adjoint representation onto
each other. The latter we construct by writing down projectors for
each of the two copies (ignoring prefactors), 
\begin{equation}
  \birdtrack{31.5ex}{PAs_box_PAa} \, , 
\end{equation}
and then seeking a non-vanishing connection inside the dashed
rectangular box. We do not have to explicitly write out such a
connection; simply notice that when we have found one then the
expression inside the grey ellipse is an invariant tensor mapping
$A$ to $A$ and thus, according to Schur's lemma, it is proportional to
$\birdtrack{7ex}{gluon_line}$. Hence, the first transition operator is
proportional to
\begin{equation}
\label{eq:PAa_to_PAs}
  \birdtrack{14ex}{PAa_to_PAs} \, .
\end{equation} 
We obtain the second transition operator by interchanging $\bullet$
and $\circ$.

\vspace{2ex}\noindent{\bf Exercise:}\nopagebreak
\begin{exc}\nopagebreak
\item Normalise the transition operator~\eqref{eq:PAa_to_PAs}.
\end{exc}
\goodbreak

\vspace{2ex}

\noindent 
{\bf $\boldsymbol{\overline{V} \otimes V \otimes A^{\otimes2}}$.}  
The quark-anti-quark pair can either be in a singlet state or in the
adjoint representation. If it is in a singlet state we need to find a
transition operator mapping the singlet within $A \otimes A$ to the
singlet within $\overline{V} \otimes V$. To this end we write down the
two projectors (ignoring prefactors) and seek a non-vanishing
connection inside the dashed box,
\begin{equation}
\label{eq:singlet_box_singlet}
  \birdtrack{33ex}{singlet_box_singlet} \, .
\end{equation}
No matter what this connection looks like, the part inside the grey
ellipse is just a number, i.e.\ the desired transition operator is
proportional to
\begin{equation}
\label{eq:singlet_to_singlet}
  \birdtrack{12ex}{singlet_to_singlet} \, .
\end{equation}

\goodbreak

\vspace{2ex}\noindent{\bf Exercise:} 
\begin{exc}
\item Find a non-vanishing way to connect the lines withing the dashed
  box in diagram~\eqref{eq:singlet_box_singlet} and evaluate the
  resulting term inside the grey ellipse.
\end{exc}
\vspace{2ex}

\goodbreak

If the quark-anti-quark pair is in the adjoint representation, then we
need to find transition operators to the two adjoint representations
within $A \otimes A$. Once more we write down the corresponding
projectors (omitting prefactors), and seek non-vanishing connections
inside the dashed boxes, 
\begin{equation}
  \birdtrack{33ex}{A_box_PAa} 
  \ , \qquad
  \birdtrack{33ex}{A_box_PAs} \, . 
\end{equation}
Once more, we do not have to find these connections explicitly, but
simply notice that the parts within the grey ellipses have to be
proportional to $\birdtrack{7ex}{gluon_line}$, i.e. the transition
operators are proportional to
\begin{equation}
\label{eq:A_to_PAas}
  \birdtrack{14ex}{A_to_PAa} 
  \qquad \text{and} \qquad
  \birdtrack{14ex}{A_to_PAs} \, .
\end{equation}
The three tensors in \eqref{eq:singlet_to_singlet} and
\eqref{eq:A_to_PAas} form an orthogonal multiplet basis for the colour
space within $\overline{V} \otimes V \otimes A^{\otimes2}$, which is
to be compared to the non-orthogonal trace
basis~\eqref{eq:trace_basis_VbarVA^2}.

\vspace{2ex}\noindent{\bf Exercise:} 
\begin{exc}
\item Normalise the basis vectors in \eqref{eq:singlet_to_singlet} and
  \eqref{eq:A_to_PAas}.
\end{exc}


\section{Further reading}
\label{sec:resources}

I give some recommendations for where to learn more about the topics
touched upon in these lectures. The list is by no means complete.
If you are interested in the history of birdtracks, I recommend
Sec.~4.9 of {\sc The Book}.

\vspace{1ex}

\noindent {\bf General introductions to and compendiums for birdtracks}
\begin{itemize}[itemsep=1ex,topsep=1ex,parsep=0mm,partopsep=0mm]
\item {\sc The Book} on birdtracks is Predrag Cvitanovi\'c's {\it
    Group Theory: Birdtracks, Lie's and Exceptional Groups}
  \cite{Cvi08}. You want to have it next to you whenever you do
  birdtrack calculations. A precursor are Cvitanovi\'c's 1984 lecture
  notes \cite{Cvi84} which already contain a lot of the material
  covered in {\sc The Book}. The 1976 paper \cite{Cvi76} provides a
  nice introduction and summary of birdtracks for $\SU(N)$ and QCD.
\item Roger Penrose developed birdtracks to be used for tensors in
  general relativity. You get a nice impression from his semi
  pop-science book {\it The Road to Reality} \cite{Pen05}. There,
  birdtrack diagrams are introduced in Secs.~\S12.8, \S13.3--\S13.9,
  \S14.3, \S14.4, \S14.6, and \S14.7 in the context of differential
  geometry and later (Secs.~\S19.2, \S19.6, \S22.12, \S26.2, \S29.5)
  used for general relativity and other topics.
\item The book {\it Diagram Techniques in Group Theory} by Geoffrey
  E.\ Stedman \cite{Ste90} also treats many aspects and applications
  of the the birdtrack method; it also contains an introductory
  section on vector algebra.
\item Yuri L.\ Dokshitzer's lecture notes {\it Perturbative QCD (and
    beyond)} \cite{DokSchSlo97} introduce birdtrack techniques using
  QCD processes as examples; some of his notational conventions differ
  slightly differ from ours.
\item If you are looking for a cheat sheet on birdtracks I suggest
  App.~A of my paper \cite{KepSjo12} with Malin Sj\"odahl.
\end{itemize}

\noindent {\bf Background}\\
Classic results on the representation theory of finite groups (such as
$S_n$) and compact Lie groups (such as $\SU(N)$), on Schur's lemma, or
on the multiplication of Young diagrams can, e.g., be found in
\cite{Ham62,Tun85,Sim96} and in a vast number of other textbooks.

\vspace{1ex}

\goodbreak
\noindent {\bf Details on the specific topics (of later sections) of
  this course}
\begin{itemize}[itemsep=1ex,topsep=1ex,parsep=0mm,partopsep=0mm]
\item Young operators (non-Hermitian) in birdtracks are discussed in
  \cite{ElvCviKen05}.
\item The main references for the construction of multiplet bases are
  \cite{KepSjo12,KepSjo14}. Hermitian Young operators are constructed
  in \cite{KepSjo14}, gluon projectors and the general rules for
  constructing multiplet bases are derived in \cite{KepSjo12}.
  \\[1ex]
  Hermitian Young operators and multiplet basis for few quarks appear
  already in \cite{CviLauSch81, Can78, Cvi08}. The multiplet basis for
  $A^{\otimes 2}\to A^{\otimes 2}$ in birdtracks is constructed in
  \cite{Cvi08} using a different method.
  \\[1ex]
  Simplification rules for a more efficient construction of Hermitian
  Young operators are derived in \cite{AlcWei17a,AlcWei17b}. The
  corresponding quark multiplet bases are discussed in
  \cite{AlcWei17c}
\item When pen and paper calculations become unwieldy, the Mathematica
  and {\tt C++} packages \cite{Sjo13,Sjo15} by Malin Sj\"odahl come in
  handy.
\item Sj\"odahl and co-workers discuss decomposition into multiplet
  bases \cite{SjoTho15} and recursion relations \cite{DuSjoTho15}. 
\end{itemize}

\noindent {\bf Typesetting.} All birdtracks in these notes where drawn
with {\tt JaxoDraw} \cite{BinThe04}.

\vspace{2ex}

\noindent {\bf Finally,} on the virtue of different notations consider
the following three cartoons\footnote{licensed under a Creative
  Commons Attribution-Noncommercial 3.0 United States License,
  \url{creativecommons.org/licenses/by-nc/3.0/us/}} from
\url{abstrusegoose.com}:%
\\[1.5ex]
\parbox{.3\textwidth}{\centering
\includegraphics[width=.3\textwidth]{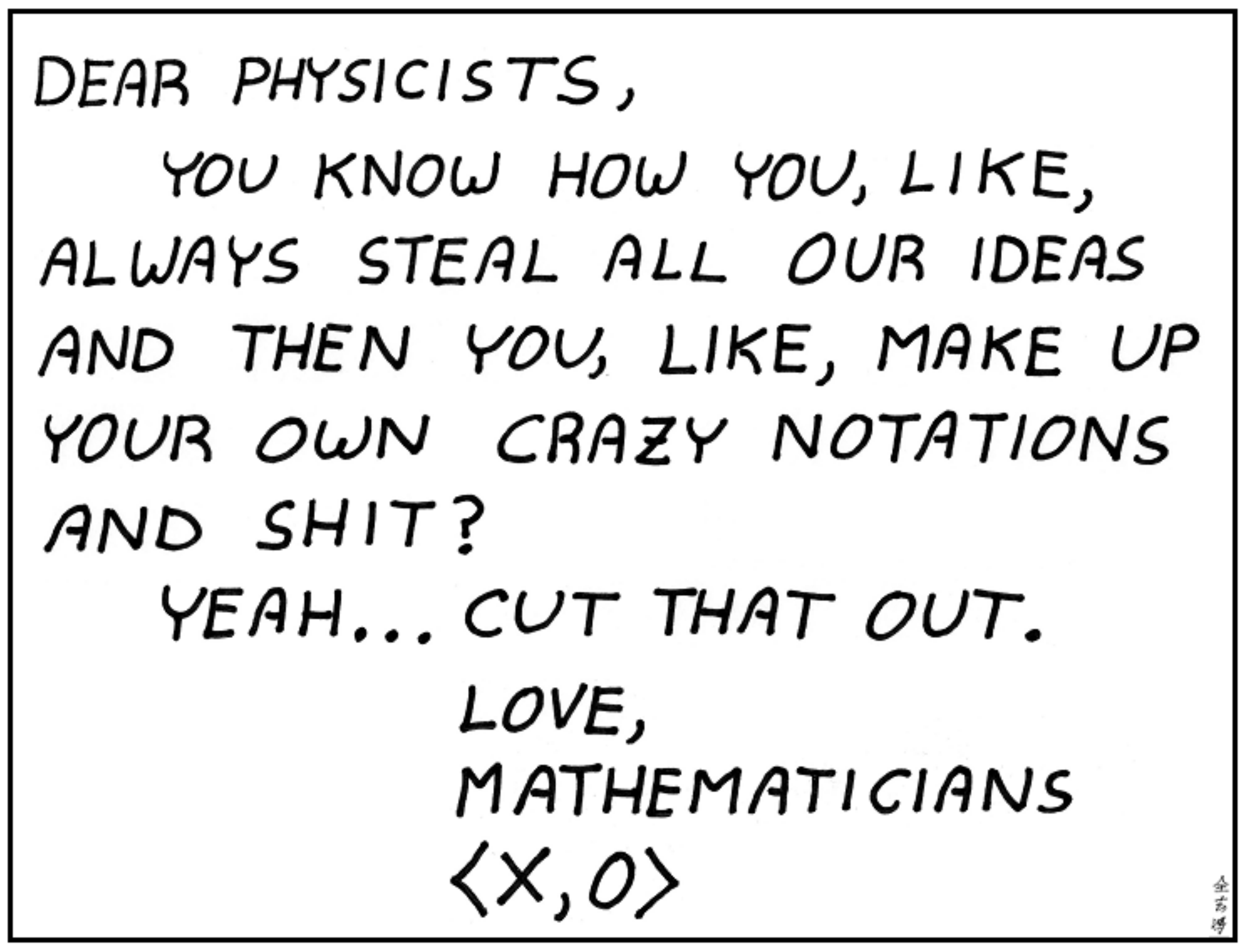}\\
\scriptsize \url{http://abstrusegoose.com/128}}
\hfill
\parbox{.3\textwidth}{\centering
\includegraphics[width=.3\textwidth]{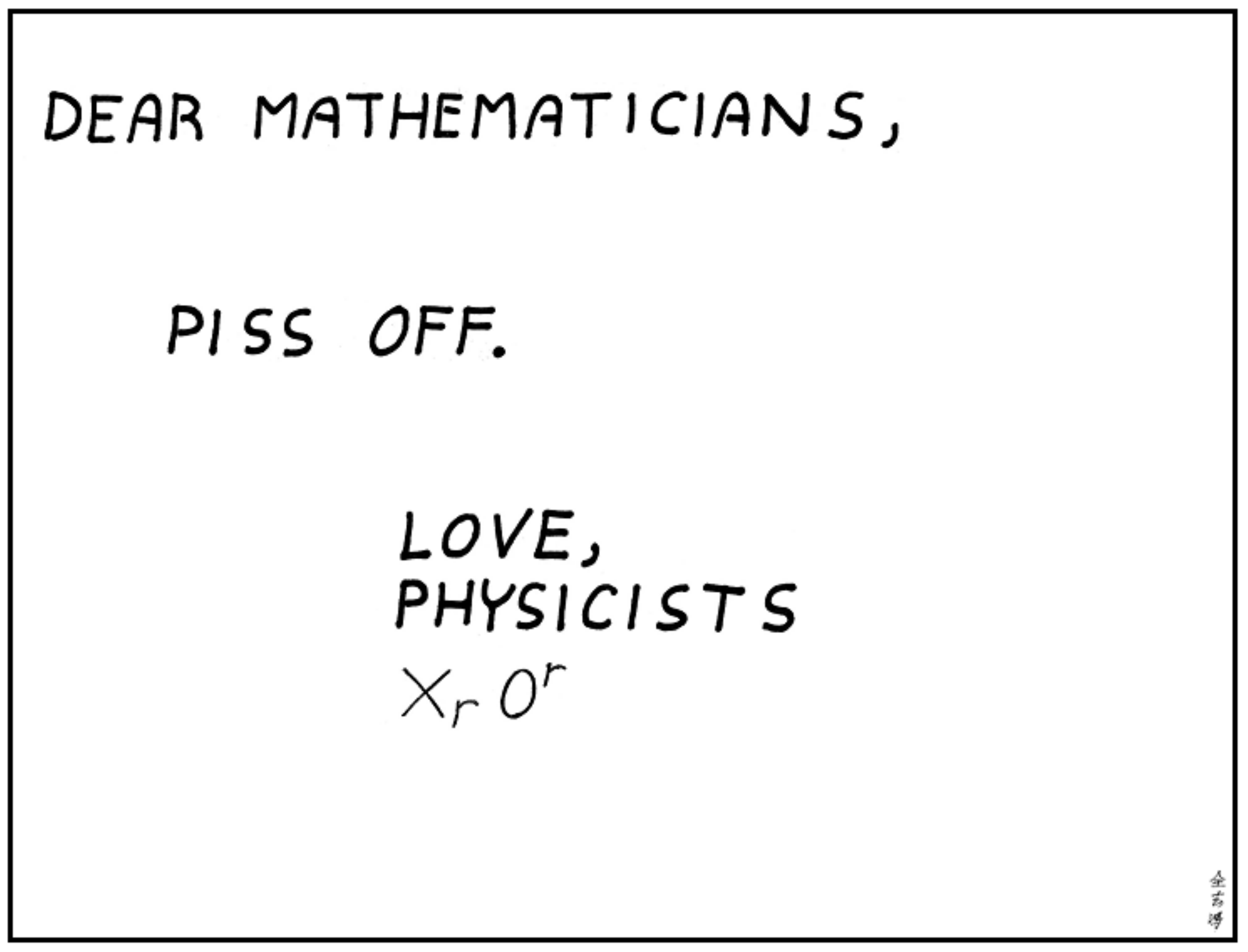}\\
\scriptsize \url{http://abstrusegoose.com/129}}
\hfill
\parbox{.3\textwidth}{\centering
\includegraphics[width=.3\textwidth]{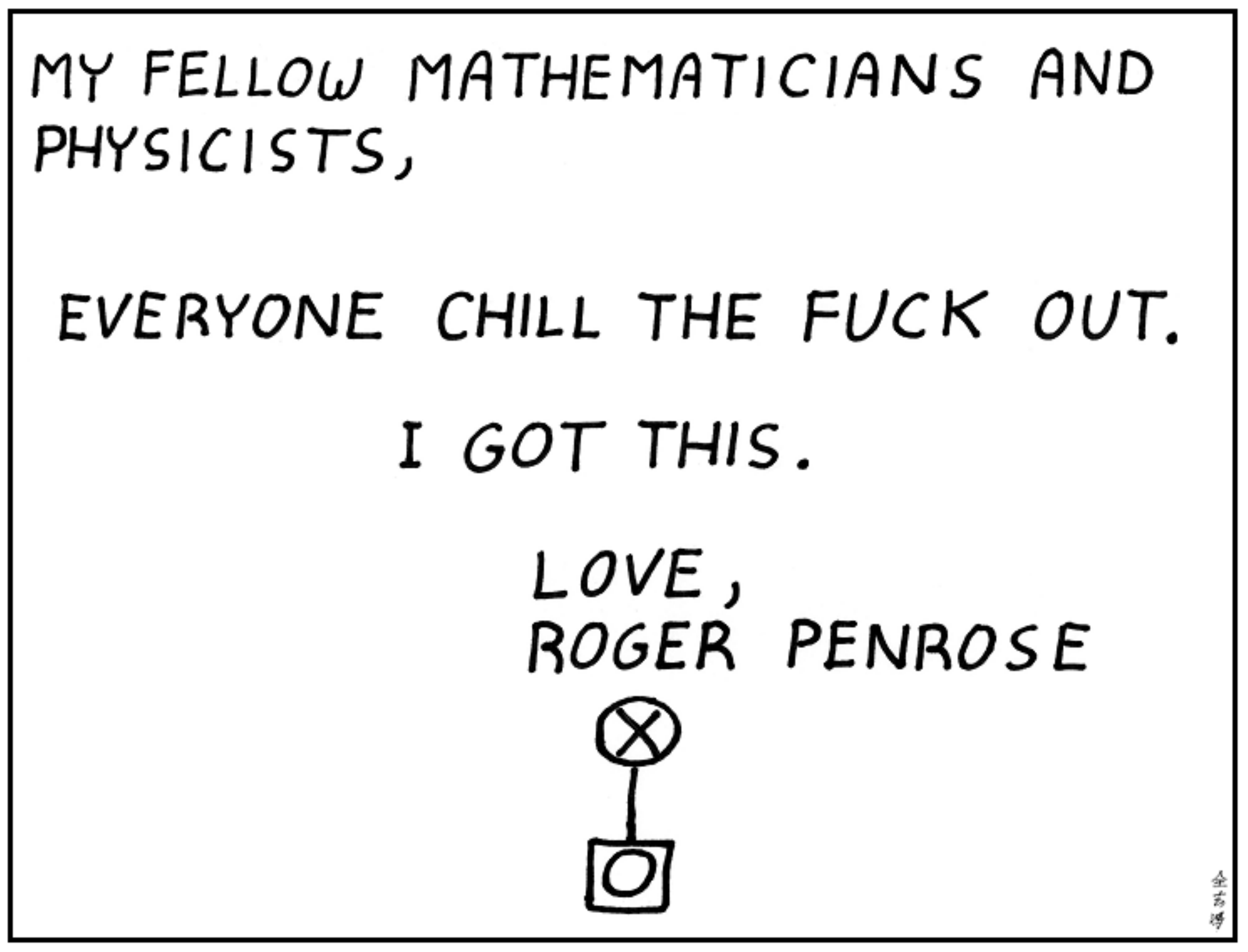}\\
\scriptsize \url{http://abstrusegoose.com/130}}
\goodbreak

 

\addcontentsline{toc}{section}{References}


\end{document}